\documentclass[reprint,amsmath,amssymb,floatfix,aps,showkeys]{revtex4-2}

\usepackage{graphicx}% Include figure files
\usepackage{dcolumn}% Align table columns on decimal point
\usepackage{bm}% bold math
\usepackage{cmap}					% поиск в PDF
\usepackage{mathtext} 				% русские буквы в формулах
\usepackage[T2A]{fontenc}			% кодировка
\usepackage[utf8]{inputenc}			% кодировка исходного текста
\usepackage[english,russian]{babel}	% локализация и переносы

\graphicspath{{images/}{images/}}  % папки с картинками

\renewcommand{\epsilon}{\ensuremath{\varepsilon}}
\renewcommand{\phi}{\ensuremath{\varphi}}
\renewcommand{\kappa}{\ensuremath{\varkappa}}

\renewcommand{\vec}[1]{\ensuremath\bm{#1}}
\renewcommand{\eqref}[1]{Eq.~(\ref{#1})}

\usepackage{hyperref}
\usepackage[usenames,dvipsnames,svgnames,table,rgb]{xcolor}
\hypersetup{				% Гиперссылки
	unicode=true,           % русские буквы в раздела PDF
	pdftitle={Заголовок},   % Заголовок
	pdfauthor={Автор},      % Автор
	pdfsubject={Тема},      % Тема
	pdfcreator={Создатель}, % Создатель
	pdfproducer={Производитель}, % Производитель
	pdfkeywords={keywords1} {key2} {key3}, % Ключевые слова
	colorlinks=true,       	% false: ссылки в рамках; true: цветные ссылки
	linkcolor=red,          % внутренние ссылки
	citecolor=black,        % на библиографию
	filecolor=black,      % на файлы
	urlcolor=black           % на URL
}

\usepackage{booktabs}
\usepackage{multirow}
\newcommand{\frc}[2]{\raisebox{2pt}{$#1$}\big/\raisebox{-3pt}{$#2$}}
\usepackage{float}

\usepackage{pstool}
 
\usepackage{adjustbox}

\usepackage{tikz}
\usetikzlibrary{shapes.geometric, arrows}

\tikzstyle{cdft} = [rectangle, rounded corners, minimum width=1cm, minimum height=1cm,text centered,text width=2.6cm, draw=black]

\tikzstyle{vfdft} = [rectangle, rounded corners, minimum width=1cm, minimum height=1cm,text centered,text width=2.6cm, draw=red]

\tikzstyle{empty} = [minimum width=1cm, minimum height=1cm,text centered,text width=1cm]

\tikzstyle{arrow} = [thick,->,>=stealth]

\usepackage[justification=centerlast]{caption}
\usepackage{subcaption}
\usepackage[section]{placeins}

\begin{document}

\selectlanguage{english}
% !TEX root = main.tex
\preprint{APS/123-QED}

\title{Bridging scales in porous media: cDFT-informed pore network modelling for fluid transport with nanoconfined phase behavior}% Force line breaks with \\

\author{Irina Nesterova}%
\email{irina.nesterova@phystech.edu}
\affiliation{%
Laboratory for Disordered Systems, Phystech School of Applied Mathematics and Computer Science, Moscow Institute of Physics and Technology, Dolgoprudny, Russia}

\author{Rustem Sirazov}
\email{sirazov@phystech.edu}
\affiliation{%
Laboratory for Disordered Systems, Phystech School of Applied Mathematics and Computer Science, Moscow Institute of Physics and Technology, Dolgoprudny, Russia}

\author{Aleksey Khlyupin}%
\email{khlyupin@phystech.edu}
\affiliation{%
Laboratory for Disordered Systems, Phystech School of Applied Mathematics and Computer Science, Moscow Institute of Physics and Technology, Dolgoprudny, Russia}

\date{\today}% It is always \today, today,

\begin{abstract}

The simulation of fluid flow in real, multiscale porous media remains challenging due to the complexity of nanoscale phenomena and the difficulty of developing upscaling methodologies. In this study, we introduce a multiscale filtration framework based on quasi-static Pore Network Modelling, incorporating the effects of pore blockage resulting from capillary condensation of fluid in the nanoporous space. To accurately predict capillary condensation in nanoconfinement, we apply classical Density Functional Theory calculations considering capillary hysteresis. The pores blocked by condensate are excluded from the fluid flow, resulting in a decrease in permeability of the porous space. Our findings demonstrate that the resulting permeability is strongly dependent on the geometry of the porous space, including pore size distribution, throat size distribution, sample size, and the particular structure of the sample, as well as thermodynamic conditions and processes, specifically pressure growth or decrease. Overall, the presented research contributes valuable insights into multiscale transport phenomena and facilitates the advancement of upscaling techniques.

\end{abstract}

%\selectlanguage{english}
%\def\tocname{Content}
%\onecolumngrid
\maketitle
%\onecolumngrid
%\tableofcontents
%\onecolumngrid
\newpage
\section{Introduction}

Fluid flow through complex multiscale porous media is a key process for numerous modern technologies, including applications in charge accumulation \cite{zhang2017multiscale, pilon2015recent}, water and air purification \cite{iliev2014multiscale, jiang2024bio, xiang2010multiscale, miller2014carbon, shao2024using}, and oil and gas recovery \cite{wang2017review, wang2020modeling, kazemi2024wettability}.
Particularly, unconventional shale reservoirs, which consist of organic and inorganic minerals, feature pores that vary in size from nanometers to micrometers \cite{ji2017pore, wang2017review, zhang2017comparisons, wu2019multiscale, shi2019modelling, jacob2020simulating, liu2021numerical, ruspini2021multiscale, zhao2022multi}. This diversity in pore scale results in significant structural heterogeneity, increasing the complexity of simulating processes occurring within multiscale structures.

To accurately describe fluid transport in such complex porous media, the development of multiscale simulation techniques is required. Pore Network Modeling (PNM) is widely recognized as a prominent tool for simulating transport phenomena with account of the pore space structure and topology \cite{blunt2001flow, blunt2002detailed, joekar2012analysis, gostick2016openpnm, cui2022pore}. It enables the analysis of flow properties, such as permeability, in relation with rock properties inside multiscale porous media \cite{jiang2013representation, mehmani2013multiscale, ruspini2021multiscale, zhang2021pore}. 
% from zhang2021: This method was used to determine the rock and flow properties of three different rock samples, such as porosity, capillary pressure, absolute permeabilities, and oil–water relative permeabilities. The pore network method was further used to determine the properties of rock matrices, such as pore size distribution, topological structure, aspect ratio, pore throat shape factor, connected porosity, total porosity, and absolute permeability.
This approach is widely used in various applications involving porous materials, including petroleum engineering \cite{yang2019pore, zhang2021anisotropic, qin2024modeling, zahasky2020pore}, energy storage and conversion \cite{gostick2007pore, zenyuk2015coupling, lombardo2019pore, sadeghi2019exploring, mckague2022extending, obliger2014pore}, chemical and biological applications \cite{ xiong2016review, lin2021simulation, sadeghnejad2021digital,  jendersie2024neuropnm}. Using a multiscale PNM approach, it was shown that the presence of microporosity significantly changed the original macropore network flow patterns \cite{wu2024multiscale}. Besides, it was adopted to reproduce grain dissolution and pore filling with clay for studying the microporosity effect on fluid filtration \cite{mehmani2014effect}. Since macroscale and nanoscale pores are considered to be in thermodynamic equilibrium, fluid properties inside them are assumed to be equal to the bulk fluid properties defined by classical thermodynamic models. However, numerous molecular simulation studies demonstrated that fluid properties in nanopores significantly differ from those observed in macropores \cite{liu2019review, wang2021molecular, sun2023review}. It was also proved experimentally by the analysis of the small-angle neutron scattering (SANS) signals for stable microemulsion inside nanoporous controlled pore glasses (CPG) \cite{dahl2024confinement}. Thus, the transport of fluids within such complex multiscale porous media exhibits behavior that deviates from classical Darcy's law due to nanoscale effects \cite{wang2015impact, wang2018pore, asai2022non}. 
 
The behavior of fluids within nanopores has been extensively studied through molecular simulations, including molecular dynamics (MD) \cite{le2015propane, wang2016molecular, wang2016fast, elola2019preferential, nan2020slip}, classical density functional theory (cDFT) \cite{ravikovitch2001density, neimark2003bridging, wu2006density, li2014phase, liu2019competitive}, and Grand Canonical Monte Carlo (GCMC) simulations \cite{coasne2004grand, liu2015adsorption, wang2019competitive, bi2019molecular}. Molecular-scale research typically focuses on fluid behavior within individual nanopores of varying geometries (e.g., slit, cylindrical, spherical) and chemical compositions (e.g., graphite, quartz, calcite), assuming thermodynamic equilibrium with the surrounding macroscale porous media. The interactions between fluid molecules and pore walls lead to the formation of a densely confined fluid exhibiting distinct thermodynamic and dynamic properties at the nanoscale. For example, it was demonstrated that the density of the ethane adsorbed layer on silica nanopores is up to 10 times larger than its density in the bulk \cite{elola2019preferential}. Other studies showed that adsorption behavior deviates with pore size and observed a tendency of heavier hydrocarbons to fill the pore \cite{ravikovitch2002density, li2014phase}. Besides, it was shown that the density and velocity profiles change depending on the material of the pore walls \cite{wang2016fast}. Therefore, the confined fluid properties depend on thermodynamic conditions, pore geometry, and material, as well as the chemical composition of the fluid, and can dramatically deviate from the bulk properties. 

A critical question arises: how can this nanoscale specificity be effectively integrated into PNM? We found two possible ways available in the literature: (i) PNM with incorporated nanoscale effects \cite{ma2014pore, zhang2015micro, chen2021pore, feng2024new}, and (ii) a combination of MD and PNM \cite{wang2017multiscale, wang2020multiscale, yu2020multiscale}, which we discuss below in detail.

The first group of studies modifies PNM to account for different effects occurring at the nanoscale. For instance, in the Refs. \cite{ma2014pore, zhang2015micro} the authors integrated different flow regimes, ranging from continuum (or viscous or Darcy), to slip, to transitional, to Knudsen diffusion, depending on the pore size and thermodynamic conditions \cite{javadpour2009nanopores, nesterova2021simulations}. Based on these flow regimes, Zhang et.al. \cite{zhang2015micro} applied an artificially generated 3D micro/nano PNM, with pore sizes in the range 100--500~nm and throat sizes in the range 1--10~nm, to study gas flow in a shale matrix. They showed that apparent permeability is strongly dependent on pore pressure in the reservoir and throat size. A similar result for gas transport in shale matrix was observed by Ma et al. \cite{ma2014pore}, who constructed a realistic 3D shale model represented by a nanoscale PNM, where about 80\% of elements were lower than 10~nm and the other lower than 50~nm. In addition to flow regimes, they also considered two van der Waals’ parameters for a non-ideal gas to study the flow of gases (methane and nitrogen) through the shale. The resulting gas permeability grows with pore size, but decreases with the gas pressure and Tangential Momentum Accommodation Coefficient (TMAC). Particularly, under normal field operational conditions (meaning temperatures close to the critical temperature of methane, which is about 190~K and pressures up to 30~MPa), the permeability of methane rises by 30\% if non-ideal gas is considered. However, these works covered only the effects of fluid transport at the nanoscale, neglecting the nanoconfined fluid phase behavior.
% Zhang 2015
% scale: micro/nano (meaning pores are micro 100-500 nm and throats are nano 1-10 nm)
% object: shale + gas
% Ma 2014
% scale: they uses realistic 3D shale model, where 80% of elements are smallet 10 nm and the others are smaller than 50 nm, i.e. nanoscale: macro (about 100 mn) + nano (about 10 nm)
% object: shale + methane or nitrogen

For that purpose, Chen et al. \cite{chen2021pore} accounted confinement-induced phase behavior to study multicomponent hydrocarbon mixture filtration in the shale rock. The authors integrated a generalized phase equilibrium model, which distinguishes adsorbed and bulk fluid phases, into PNM to represent the multiscale fluid behavior. They considered three artificially constructed PNM: one two-scale with large pore 100~nm and smaller pores 10~nm, and two multiscale with pores range in between 5.5--100~nm, but with different average pore sizes of 12.2~nm and 22.7~nm, correspondingly. It was shown that the phase behavior in the PNM, controlled by the multiscale pore structure, significantly deviates from that in a single nanopore. Besides, due to capillary trapping of the liquid phase and competitive adsorption on the pore wall, heavier components tend to reside in smaller pores and suppress the bubble point pressure therein. Overall, we can conclude that the specificity of fluid behavior at the nanoscale, both dynamical and thermodynamic, should be accounted for to accurately describe fluid filtration through multiscale porous media. However, the porous structures considered in these works were simplified, meaning artificially generated or considering only one scale of pores. 
% scale: multiscale: macro (about 100 mn) + nano (about 10 nm), multiscale in range from 5.5 nm to 100 nm
% object: shale + multicomponents hydrocarbon mixture

%Works on multi-scale PNM + nanoeffect:

Recently, the specificity of flow dynamics at the nanoscale was integrated into a realistic multiscale PNM. Feng et al. \cite{feng2024new} proposed a dual-scale PNM to study gas flow in shale. In this work, the information of porous structure on microscale was taken from N$_2$ adsorption or mercury intrusions porosimetry, while the nanoscale information is given by Focused Ion Beam Scanning Electron Microscope (FIB-SEM). This model accounts for different types of pores: (i) organic pores, (ii) inorganic pores, and (iii) inorganic pores with clay. Despite the accurate representation of porous structure, they considered the swelling behavior of clay upon hydration and various flow mechanisms, including Knudsen diffusion for inorganic pores and a combination of Knudsen and surface diffusion for organic pores. The analysis reveals that permeability decreases with increasing maximum organic pore radius, roughness, clay content, water film thickness, and tortuosity. Conversely, permeability increases with higher total organic carbon (TOC) content and improved pore connectivity. However, the works considered in the first group used a continuum-based or analytical model to describe the specificity of fluid behavior at the nanoscale without validation on the results of molecular scale approaches.
%  scale: multiscale: Micro (N2 ads or MIP mercury intrusion porosimetry) + nano (FIB-SEM)
%  object: shale + gas
%In another work, Qin et al.\cite{qin2024modeling} studied 2 phase flow of water/oil system in the low permeability sandstone. For that purpose, the authors developed a dual-porosity PNM, where the macroscale and microscale information is taken from the micro CT and the information about microporosity is given through ???. They incorporated the effects of boundary slip and interphase drag through a free surface model to determine the water-oil relative permeability in low-permeability samples. This model reveals that microporosity causes a decrease in the residual oil saturation and negatively correlates with invading oil volume fraction due to strong capillary resistance. 
%  scale: multiscale: macro (from microCT) + micro (??? some image proccessing)
%  object: low permeability sandstone + 2 phase flow water-oil

\begin{figure*}
    \centering
    \includegraphics[width=1\linewidth]{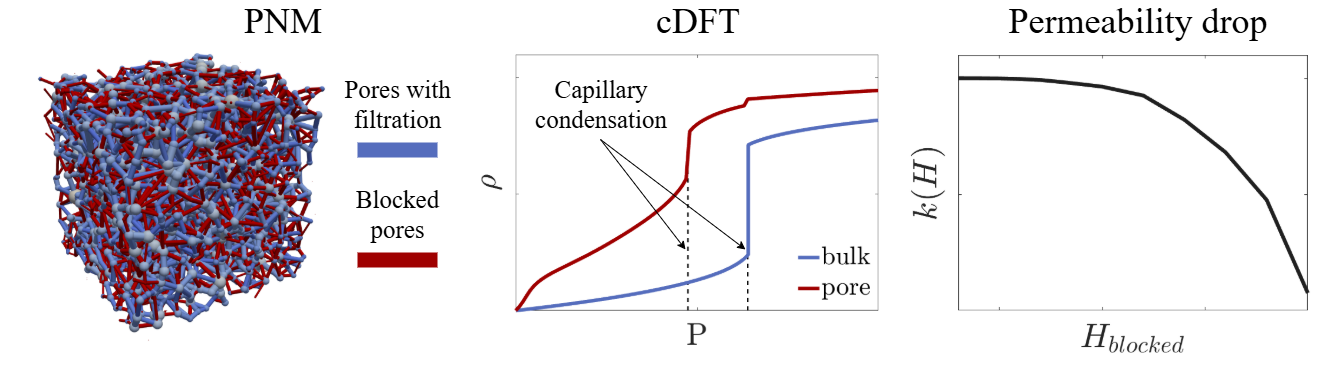}
    \caption{Schematic illustration of the presented multiscale methodology, combining PNM simulation of fluid filtration with cDFT calculation of nanoconfined fluid phase behavior. Confinement conditions promote the formation of the condensate phase inside pores at lower pressures than in the bulk, which can lead to pore blockage by a condensate in the case of low capillary number and result in a permeability drop for gas flow.}
    \label{fig:mol_scale_prop}
\end{figure*}

For a more accurate description of fluid flow at the nanoscale, in the second group of studies, MD simulations were combined with PNM. Regarding this approach, Wang S. et al. \cite{wang2017multiscale, wang2020multiscale} performed MD simulations of methane flow through the pores with different mineralogy to build modified Hagen-Poiseuille (HP) equations that describe the integrated flow for each pore. Then, these modified HP equations are used in the multiscale PNM with individual pore size distributions of organic and inorganic pores, constructing the shale matrix. As a result, flow intensification was observed only in the organic pores, while in the calcite and montmorillonite pores, the deviation from classical Poiseuille flow was insignificant. The authors concluded that permeability mainly depends on the pore size distribution (PSD) of inorganic pores, since it has larger pore diameters compared with the organic matrix, and that the permeability decreases for higher values of the TOC component. In another study, Yu. H. et al. \cite{yu2020multiscale} performed MD simulation to build a unified gas transport model that was further incorporated into the PNM approach to describe fluid flow behavior in the multiscale porous media. First, the authors perform MD simulations of shale gas transport in a single pore with different pore wall materials (calcite, quartz, montmorillonite, and organic) to describe flow on the molecular scale. Then, they analyzed slip flow behavior, which was found to be different for organic and inorganic pores, meaning a significant slippage effect observed only for organic pores. After using continuum flow theory and gas-surface dynamics theory, they proposed a unified gas transport model to describe the gas transport in porous media. Finally, they perform PNM calculations with 3 scales: nanopores less than 10~nm, micropores with sizes between 10 and 100~nm, and macropores larger than 100~nm, using the unified gas transport model. The results showed that the contribution of organic pores to shale matrix permeability is higher than that of inorganic pores, even for inorganic-rich shale matrix. The authors concluded that the gas transport of the shale matrix is mainly determined by the flow behavior in the organic pores. It was also found that the permeability decrease as the organic pores rate decreases and is influenced by the pore size and system pressure. Considering these two works, it is worth noting that, by obtaining similar flow behavior in the nanopores (both organic and inorganic), the authors come to controversial conclusions about the role of organic and inorganic content on the permeability. Moreover, the consideration of nanoscale flow behavior in the Refs. \cite{wang2017multiscale, wang2020multiscale} lead to the flow intensification, i.e., permeability increase, while in the Ref. \cite{yu2020multiscale} permeability reduction is observed. The core of this difference is in the way how nanoscale data is integrated into PNM. Obvious that nowadays there is still no agreement about the proper way for that, but the application of the unified gas transport model seems to be more accurate for the description of hard-to-recover shale reservoirs.

Since there are several works that try to integrate the molecular scale specificity of fluid transport into PNM, we have not found such works with an accurate description of the nanoconfined phase behavior from molecular scale approaches. So we decided to cover this gap in this study. To address this issue, we chose cDFT as the approach for molecular scale description of nanoconfined phase behavior. The cDFT allows us to calculate the properties of fluid state in the nanopores, less computationally expensive compared to MD calculation, especially for the larger pores (about 50~nm) and for the system of pores connected with the bulk, described by grand canonical ensemble (GCE). Besides, in our previous work \cite{vaganova2022linking}, we have shown that the equilibrium densities of fluid in the nanopores obtained using these two methods are in good agreement. Thus, we present a new multiscale methodology to simulate fluid filtration with account for nanoconfined fluid phase behavior using cDFT calculations. 

% about our focuses: pore blockage

In this study, we focus on the effects of pore blockage induced by capillary condensation in the nanopores. Due to the impact of pore walls, the PVT properties of fluids in the nanopores are changed, i.e., capillary condensation occurs at lower pressure values than in the bulk \cite{liu2019review, le2015propane, ravikovitch2001density, neimark2003bridging, li2014phase, bi2019molecular}. When capillary condensation occurs, and a pore is filled with a liquid phase, if the local pressure gradient is lower than the capillary pressure, then the gas flow through the pore is blocked \cite{fang1996phenomenological, li2000phenomenological, santos2020pore, reis2021pore, hosseinzadegan2023review}. Several studies demonstrated a reduction of gas relative permeability with growth of condensate saturation \cite{li2000phenomenological, santos2020pore, bustos2003pore,reis2020pore, reis2021pore}. It was also shown that gas-condensate flow at the capillary flow regime is governed by pore space morphology and connectivity \cite{wang2000pore, bustos2003pore, hosseinzadegan2023review}. However, they considered condensate formation in terms of gas-condensate flow or retrograde gas flow, meaning microscale of porous structures and the consideration of bulk fluid properties, rather than nanoconfined phase behavior. We decided to take this idea of pore blockage by a condensate phase considered in these studies and apply it to a nanoscale porous structure with accurate fluid phase behavior in confinement conditions.

% about our focuses: capilary hysteresis

Despite pore blockage during capillary condensation, we also consider the capillary hysteresis, meaning that the pressure value of capillary condensation upon pressure growth and capillary evaporation upon pressure decrease are different. To learn more about this phenomenon, we refer the readers to the review \cite{barsotti2016review}, where theoretical and experimental techniques used to describe capillary condensation, focusing on hydrocarbon systems, are presented. The capillary hysteresis phenomenon is quite diversified and depends on a set of properties, like length of fluid molecule chain, pore space morphology and connectivity, deformation of solid adsorbent, etc. \cite{barsotti2020capillary}. The capillary hysteresis, together with pore blockage by a formed condensate, leads to a similar permeability hysteresis of the considered porous space. This phenomenon was studied by Mehmani et al. \cite{mehmani2014application} in heterogeneous nanoscale pore network models filled with nitrogen. They have shown that pore network topology impacts the behavior of capillary pressure curves along with sorption and permeability hysteresis. Additionally, it was found that the hysteresis loop grows with reducing connectivity, however, these effects become less significant for bigger pore sizes. However, in this study for the description of capillary condensation the authors use the Kelvin equation, which can not capture the molecular scale specificity of confined fluid behavior and nanoscale material properties.

% about our work

In this paper, we simulate fluid filtration with nanoconfined phase behavior and pore blockage, combining the PNM method with cDFT calculations, by a scheme illustrated in Fig. \ref{fig:mol_scale_prop}. As an object for our study we consider carbon dioxide, since it is promising in terms of enhanced oil recovery (EOR) method and for carbon capture and storage (CCUS) applications \cite{blunt1993carbon, alvarado2010enhanced, sheng2015enhanced, tapia2018review}. The behavior of carbon dioxide is studied inside organic nanoporous structure, which is commonly used as a representation for shale matrix, for example in Refs. \cite{ma2014pore, chen2021pore}. Besides, the consideration of organic content is intriguing because of obtained disagreement of its role on the permeability obtained by the combination of PNM and MD approaches \cite{wang2017multiscale, wang2020multiscale, yu2020multiscale}. To accurately represent fluid PVT in the nanopores, the conditions of capillary condensation and capillary evaporation in the pores are calculated using the cDFT approach. Then, the obtained conditions for capillary hysteresis are given as the input in the PNM filtration calculations. We perform PNM calculations of samples permeability with account for pore blockage by a condensate, depending on the pore size and thermodynamic conditions. Additionally, the sample size, structural correlations, and size distributions of pores and throats are varied to investigate how they will influence the permeability of the sample. The results show a decrease in permeability of the sample if capillary condensation occurs in the pores and throats. Besides, the permeability hysteresis is observed due to the capillary hysteresis, meaning the permeability of the sample depends on thermodynamic processes, i.e., pressure growth or decrease. Moreover, the permeability drop is controlled by the structural characteristics of the sample, such as PSD and the size of the sample. Important findings of the presented study are that PSD is not a sufficient structural characteristic to predict the behavior of fluid filtration, and the size of the sample should be large enough to show a representative result. Thus, the results of this study provide important insights for multiscale fluid filtration. 

% !TEX root = main.tex
\section{Methods}

The multiscale filtration framework presented in this study integrates PNM with cDFT to analyze fluid filtration with the nanoconfined fluid phase behavior. It is essential to recognize that the PVT characteristics of fluids within nanopores diverge from those observed in the bulk conditions. Consequently, the phases of the fluid within macropores and nanopores can be different; for instance, while the fluid in macropores is in a vapor phase, the fluid within connected nanopores can experience capillary condensation, resulting in a liquid phase. Accounting for different phase behavior inside nano and micropores, we perform quasi-static PNM calculations under the assumption of low capillary number, meaning that capillary forces prevail over viscous ones. Thus, we assume that when the fluid inside the nanopore is in the liquid phase, high capillary forces can disrupt fluid flow, and this pore is not included for fluid filtration in the pore network. As a result, the account of nanoconfined fluid phase behavior leads to a reduction in the permeability of the sample. In the following sections, we provide a detailed description of the PNM algorithm and the cDFT calculations employed in this study, elucidating how these two approaches are combined. The overall pipeline can be described as follows: firstly, fluid PVTs in the nanopores are obtained using cDFT calculations; secondly, the permeability curves of the pore network sample are calculated with account of capillary condensation according to nanoconfined fluid PVT.

\subsection{Classical Density Functional Theory}

Classical Density Functional Theory is the rigorous method based on statistical mechanics principles, which is widely used to study fluid behavior in confinement \cite{ravikovitch2001density, neimark2003bridging, wu2006density, li2014phase, liu2019competitive}. It enables to bridge molecular scale fluid behavior with macroscopic fluid properties in the bulk using less computational requirements than molecular simulations. Recently, we have applied cDFT for accurate fluid mixture characterization \cite{nesterova2022adaptive}, developed variation free cDFT approach to speed up calculations \cite{kanygin2022variation}, and proposed linking MD-cDFT approach to optimize computationally expensive simulation for the open system of nanopore connected with bulk \cite{vaganova2022linking}. Moreover, the cDFT approach has a variety of important extension, such as Statistical Associating Fluid Theory (SAFT) for chain molecules \cite{chapman1988phase}, Quenched Solid Density Functional Theory (QSDFT) \cite{ravikovitch2006density}, and Random Surface Density Functional Theory (RSDFT) \cite{khlyupin2017random, aslyamov2017density,aslyamov2019theoretical} to account for geometrical heterogeneity of rough surfaces, Random Surface Statistical Associating Fluid Theory (RS–SAFT) combining the contribution of rough surfaces and complex liquids \cite{aslyamov2019random}, and Hydrodynamic Density Functional Theory (H-DFT) including driving forces for molecular diffusion of inhomogeneous systems \cite{stierle2021hydrodynamic}. 

To account for the nanoconfined phase behavior of fluid in the nanoporous samples, we perform classical DFT calculations for the whole range of pore widths of the considered samples. The behavior of the fluid molecules in the nanopores connected with macropores is described by the grand canonical ensemble with constant parameters $(T,V,\mu)$. The free energy of this system is described by the Omega potential $\Omega$ that is formulated as a functional of the density distribution function $\rho(\vec{r})$:
\begin{equation}\label{eq:Omega}
\Omega\left[\rho\left(\vec{r}\right)\right] =   F\left[\rho\left(\vec{r}\right)\right]
    + \int d\vec{r} \rho\left(\vec{r}\right)\left(V^{ext}\left(\vec{r}\right)-\mu\right),
\end{equation}
where $F\left[\rho\left(\vec{r}\right)\right]$ is the intrinsic Helmholtz free energy,  $V^{ext}$ is the external potential, and $\mu$ is the chemical potential.

The Helmholtz free energy functional $F\left[\rho\left(\vec{r}\right)\right]$ is the sum of terms corresponding to each type of intermolecular interaction. We consider molecular repulsion in terms of hard-sphere interactions and molecular attraction in addition to the ideal term: 
\begin{equation} \label{eq:helm_free_energy}
    F\left[\rho\left(\vec{r}\right)\right] =  F^{id}\left[\rho\left(\vec{r}\right)\right] + F^{hs}\left[\rho\left(\vec{r}\right)\right] + F^{att}\left[\rho\left(\vec{r}\right)\right]
\end{equation}
The external potential $V^{ext}$ represents the interactions between the fluid molecules and the pore walls. The expressions for the external potential $V^{ext}$ and chemical potential $\mu$  are given in Appendix \ref{sec:Appendix_DFT}.

According to the condition of thermodynamic equilibrium, $\Omega$ potential turns to the minimum at the equilibrium: 
\begin{equation}
    \frac{\delta \Omega \left[\rho\left(\vec{r}\right)\right]}{\delta \rho \left(\vec{r}\right)} = \frac{\delta F \left[\rho\left(\vec{r}\right)\right]}{\delta \rho \left(\vec{r}\right)} + V^{ext}\left(\vec{r}\right)-\mu = 0
\end{equation}

Then, the density distribution function $\rho\left(\vec{r}\right)$ is expressed as:
\begin{align} \label{eq:density}
    \tilde\rho\left(\vec{r}\right) = \rho^{bulk}\exp
    \bigg\lbrace -\frac{1}{k_BT}\bigg(
    &\dfrac{\delta F\left[\rho\left(\vec{r}\right)\right]}{\delta \rho\left(\vec{r}\right)} \nonumber \\
    & + V^{ext}\left(\vec{r}\right) - \mu^{ex}\bigg)\bigg\rbrace,
\end{align}
where $\rho^{bulk}$ is the density in the bulk, $k_B$ is the Boltzmann constant, $T$ is the system temperature, $\mu^{ex}$ is the excess chemical potential, which equals $\mu^{ex} = \mu - \mu^{id}$. 

This equation is solved by the Picard iterations, where the next iteration step is the sum of the previous step and the current solution from Eq. \ref{eq:density}, mixed with parameter $\alpha \in [0, 1]$. For the first iteration step $\rho^0 = \rho^{bulk}$.
\begin{equation}
    \rho^{j+1} = \left(1-\alpha\right)\rho^{j} + \alpha\tilde{\rho}^j
\end{equation}

\begin{figure}
    \centering
    \includegraphics[width=1\linewidth]{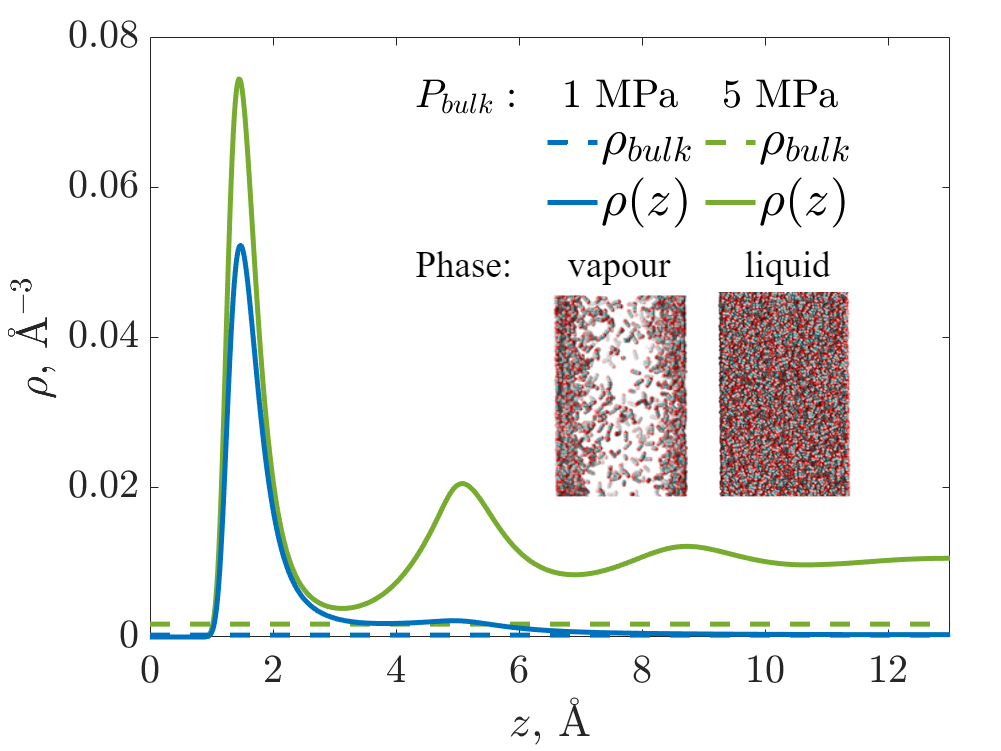}
    \caption{Density profiles in the pore H = 3~nm (solid lines) comparing to bulk density (dashed lines) for $CO_2$ at T = 298~K and P = 1 and 5~MPa.}
    \label{fig:dens_profiles_pore}
\end{figure}

To describe a particular fluid at a certain temperature, we need to set the Lennard--Jones parameters that describe intermolecular interactions between fluid molecules. These parameters are determined by fitting the cDFT EoS on the isothermal PVT data at the required temperature \cite{nesterova2022adaptive}. For calculations of carbon dioxide phase behavior in the nanopores, we use the values of Lennard--Jones parameters given in Table \ref{tab:dft_params}, which were estimated by fitting cDFT EoS on the data from the NIST Chemistry WebBook \cite{span1996new}. For the description of fluid-wall interactions, we consider organic (carbon) pore walls (for more details, see Appendix \ref{sec:Appendix_DFT}).

\renewcommand{\arraystretch}{1.1} %% increase table row spacing
\renewcommand{\tabcolsep}{0.2cm} %% increase table column spacing
\begin{table}[!h]
\caption{Intermolecular interaction parameters for $CO_2$.}
\label{tab:dft_params}
\begin{ruledtabular}
\begin{tabular}{ccc}

\multirow{2}{*}{Temperature, K} &
  \multirow{2}{*}{$\frc{\epsilon}{k_B},$~K} &
  \multirow{2}{*}{$\sigma,$ \AA}\\
            &                &                \\
                          \midrule
\multicolumn{1}{l}{273}  & \multicolumn{1}{c}{231.04}          & 3.610\\ 
\multicolumn{1}{l}{298}  & \multicolumn{1}{c}{226.39}          & 3.597\\ 
\end{tabular}
\end{ruledtabular}
\end{table}

In Fig. \ref{fig:dens_profiles_pore}, the deviation of the density of the nanoconfined inhomogeneous fluid from the bulk density is illustrated. Here, the results of the cDFT calculation for carbon dioxide in the slit-like pore with the width H = 3~nm at T = 298~K and two pressure values, P = 1~MPa and P = 5~MPa, are presented. One can see that the density profile in the pore has an oscillating character, and the local values of density can be much higher than the bulk value. Actually, the values of fluid density in the nanopore coincide with the bulk density in the center of the pore if no capillary condensation occurs, for example, in the case of P = 1~MPa. At P = 5~MPa, we observe a capillary condensation that occurs in the pore earlier than at the bulk conditions. The pressure of liquid-gas phase transition in the bulk for carbon dioxide at T = 298~K equals P$_{ph.tr.}^{bulk}$ = 6.4121~MPa. However, in the nanopore H = 3~nm, it happens at P = 3.1~MPa both during pressure growth and decrease. In the insets, for clarity, we show MD snapshots for vapor and liquid states of carbon dioxide in the pore at these thermodynamic conditions \cite{vaganova2022linking}.

We perform cDFT calculations of fluid density distribution function in the nanopores with widths from the pore and throat size distributions of the sample during pressure increase and decrease. Then, we calculate the average density of fluid in the pores for each pressure and pore width, thereby obtaining the effective fluid PVT in the pore for two cases: pressure increase and decrease. When capillary condensation occurs, the average density has a sharp jump up in the case of pressure increase. Similarly, during pressure decrease, the average fluid density sharply drops, illustrating capillary evaporation inside the pore. The resulting PVT curves of these two cases are different because capillary condensation and capillary evaporation occur at different values of pressure, i.e., so-called capillary hysteresis. The resulting fluid PVT in the nanopores is further used during PNM calculation, where the pores with condensate are blocked during gas phase filtration.

\subsection{Pore Network Model}\label{sec:PNM}

In the Pore Network Modelling, pore space is described as a graph. Nodes of the graph represent pore bodies, and edges are representations of interconnections between pore bodies called throats. This representation is made to reduce the dimensionality of the problem. Instead of complex representations of pores and throats in voxel images or polygonal meshes, PNMs use simple low-level structures, for example, a classic square, triangle, or circle \cite{blunt2002detailed}. In this section, we will describe the pipeline of computing petrophysical properties used in our study. 

Taking into account the problem statement, we chose a quasistatic, isothermal pore network model. Capillary numbers tending to zero allowed us to stick with a single-phase model with capillary condensation taken into account by modification of the pore network.

Estimation of transport properties starts with the flow rate calculation by solving a system of linear equations, considering that the condition of mass conservation in every pore is met:
\begin{eqnarray}
\sum_{j} q_{ij}=0
\label{eq: qij =0}
\end{eqnarray}

The flow rate $q_{ij}$ between two pores $i, j$ is defined as follows:
\begin{eqnarray}
q_{ij}=\frac{g_{ij}}{L_{ij}}\Delta P_{ij}
\label{eq: qij}
\end{eqnarray}
where $g_{ij}$ is the conductance between pores $i$ and $j$, $L_{ij}$ is the length between pore centers, and $\Delta P_{ij}$ is the pressure difference between the pores. For a pore network element, the conductance given by Poiseuille's law can be adopted:
\begin{eqnarray}
g=k\frac{A^2G}{2\mu}
\label{eq: g}
\end{eqnarray}
with $k$ being $1/2$, $3/5$, $0.5623$ for circle, triangle, and square cross-section, respectively \cite{Oren_1998}.  Pore-to-pore conductance then can be calculated as a harmonic mean of pores and throat conductances:
\begin{eqnarray}
\frac{L_{ij}}{g_{ij}}=\frac{L_i}{g_{p,i}}+\frac{L_t}{g_{t}}+\frac{L_j}{g_{p,j}}
\label{eq: g_ij_mean}
\end{eqnarray}
where $g_{p,*}$ is pore conductance, $g_{t}$ is throat conductance, $L_{t}$ is throat length,  $L_{i},L_{j}$ are the distance between pore center and pore-throat interface.

Single-phase permeability of the network, then, is found from Darcy's law:
\begin{eqnarray}
K=\frac{\mu qL}{A\Delta P}
\label{eq: darcy}
\end{eqnarray}
where $\mu$ is fluid viscosity, $A$ is the cross-sectional area of the sample, $L$ is the length of the sample, $\Delta P$ is the pressure drop on the sample, and $q$ is the total flow rate, which is calculated in the perpendicular to the considered flow direction cross-section. In our case, it is calculated as the mean of the input and output flow of the sample.

To incorporate capillary condensation in pores and throats due to the confinement effect, a few changes to the described above pore network algorithm were made. Firstly, we presume that if fluid in the pore network element is condensed, it prevents the flow of gaseous fluid through the element. An argument can be made that in the case of widespread condensation, there can be two-phase flow presented. However, in this paper, we work on the premise of mostly gaseous fluid flow with a part of the elements subjected to condensation. Secondly, we disable pore network elements in which condensation has occurred. We achieve this by removing corresponding elements from the matrix defined in  Eq. \ref{eq: qij =0} and Eq. \ref{eq: qij}. Then, we calculate the single-phase permeability of the truncated pore network for all pressure points during isotherms with capillary hysteresis consideration. Different pressure values correspond to different sizes of pore network elements where condensation occurred. The result of the calculation is then compared to the single-phase permeability of the pore network without the condensation effect. 

\subsection{Sample generation}

\begin{figure}
    \centering
    \includegraphics[width=1\linewidth]{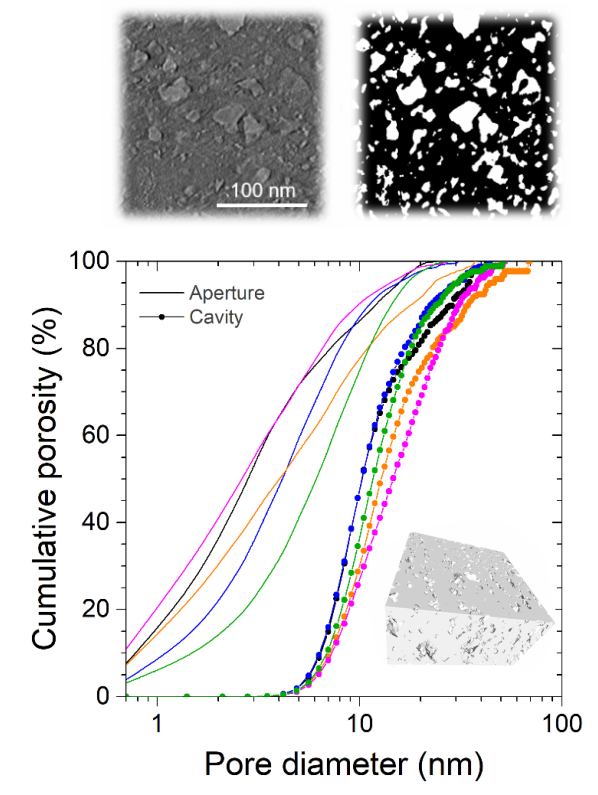}
    \caption{Cumulative histograms of the aperture (solid line) and cavity (line with symbols) size distribution of the VM samples along with electron tomography snapshot, segmentation image, and 3D reconstruction. Adopted from Ref. \cite{berthonneau2021nanoscale}. Copyright 2021 American Chemical Society.}
    \label{fig:vm}
\end{figure}

The present study focuses specifically on nanoscale porous structures. Comprehensive characterization of the pore space remains constrained by the inherent limitations of available techniques, including the resolution threshold of computed tomography \cite{Sharma2011}, surface-restricted information from electron and atomic-force microscopy \cite{chen2021pore}, and the trade-offs between sample size and resolution in electron tomography and FIB-SEM \cite{Loucks2012, Ruspini2016}. Furthermore, achieving both representativity and reasonable computational accuracy in digital samples is challenging due to the possibility of higher-order heterogeneity \cite{Zhang2000, chen2021pore}. Heterogeneity on larger scales motivates the development of pore space reconstruction methods \cite{jiang2013representation, Bultreys2015, Kulygin2024}.

Consequently, we rely on artificial samples derived from the procedural generation of a pore network within a cubic domain. To connect the synthetic samples to a real formation, a subset of the designs employs a throat and pore size distribution derived from the Vaca-Muerta Formation \cite{berthonneau2021nanoscale}, shown in Fig. \ref{fig:vm}. To begin with, we specify the total number of pores. This number sets the uniform grid resolution by determining the count of lattice nodes along each of the three axes. The node count, in turn, defines the regular center-to-center spacing between adjacent pore bodies. Within this cubic framework, pore elements are placed while throats are instantiated to connect each pore to its 6 nearest neighbors. This approach leads to a uniform and deterministic cubic regular lattice. Then, procedural variation is introduced by sampling pore and throat radii from predetermined statistical distributions. Throat sizes are drawn randomly from a chosen throat size distribution (TSD); two TSDs are considered: uniform (uni) and a distribution derived from Vaca Muerta Formation core samples (VM) \cite{berthonneau2021nanoscale}. Both distributions are truncated at 60~nm, since pores larger than 60~nm constitute less than $10^{-4}$ of the VM sample \cite{berthonneau2021nanoscale}, and this truncation is applied uniformly for consistency. Pore sizes are then assigned with the constraint that each pore is no smaller than its connected throats. 

Two distinct methods are used for the spatial arrangement of pores inside a predefined cubic lattice. Random placement, where the element sizes (pores and throats) are placed at a lattice without any spatial correlation. Space‑correlated placement, where a 3D stochastic Gaussian random field is generated by open-source code from GitHub \cite{3DGauss}, parameterized by the average power $P_0$, the parameter $\zeta \in [0,1]$, which defines the dilatational or solenoidal nature of the field, and the maximum wave-number $k_1$. The initially random element size distributions (PSD and TSD) values are then sorted according to the field amplitude, which introduces a prescribed spatial correlation of pore sizes.
    
Four distinct series of artificial pore network samples were generated for this study. Parameters of generation are provided as follows: 

\begin{itemize}
    \item Uniform throat size distribution with random elements placement. The linear size of the bounding box is 500~{\textmu}m with 125000 pores in the sample. Sample, referred to as 'uni-500', was used in the experiment series given in sections \ref{sec:perm_drop}, \ref{sec:perm_T}, and \ref{sec:perm_psd}.
    \item Throat and pore size distributions based on core samples from the Vaca Muerta Formation \cite{berthonneau2021nanoscale} with random elemenst placement. The linear size of the bounding box is 500~{\textmu}m with 125000 pores in the sample. Sample, referred to as 'vm-500', was in the experiment series given in sections \ref{sec:perm_psd}.
    \item Uniform throat size distribution with random elements placement.  The linear sizes of the bounding box are 240, 480, 720, 960~{\textmu}m with 512, 4096, 13824, 32768 pores, correspondingly. Samples, referred to as 'uni-240', 'uni-480','uni-720','uni-960', were used in the experiment series given in section \ref{sec:perm_size}.
    \item Uniform throat size distribution with space-correlated elements placement. The linear size of the bounding box is 512~{\textmu}m with 32768 pores in the sample. Six different samples with various Gaussian field parameters, referred to as Sample \#1-\#6, were used in the experiment series given in sections \ref{sec:perm_struct}.
\end{itemize}
%\input{sections/materials}
% !TEX root = main.tex
\newpage
\section{Results}

To account for the fluid PVT in confinement, we perform cDFT calculations of average density for carbon dioxide in the range of pores H = 3--60~nm at fixed temperatures T = 273~K and T = 298~K and different pressure values up to 10~MPa. Then, we integrated the results on nanoconfined phase behavior of carbon dioxide into PNM calculations and evaluate permeability of different samples, varying sample size, element size distributions, and particular porous space structure, accounting for pore blockage by a condensate (liquid) phase. In the Results \ref{sec:perm_drop}, for clarity, we consider a case study for the permeability drop for carbon dioxide at T = 298~K inside the 'uni-500' sample. Then, in the Results \ref{sec:perm_T}, we analyze capillary and permeability hysteresis at different temperatures T = 273~K and T = 298~K inside the 'uni-500' sample. In the following subsections, Results \ref{sec:perm_psd}, \ref{sec:perm_size}, and \ref{sec:perm_struct}, we investigate the role of sample geometry characteristics, such as size, element size distributions, and particular structure, on the permeability drop, considering nanoconfined phase behavior for carbon dioxide at T = 273~K. Results \ref{sec:perm_size} also used for representative elementary volume investigation \cite{zubov2024search} for the samples with uniform TSD with random elements placement. The same investigations were made for the samples with elements size distributions based on Vaca Muerta Formation core samples \cite{berthonneau2021nanoscale} with random element placements and for the samples with uniform TSD with space-correlated elements placements before conducting calculations.

\subsection{Permeability drop and hysteresis}\label{sec:perm_drop}

\begin{figure}
    \centering
    \includegraphics[width=1\linewidth]{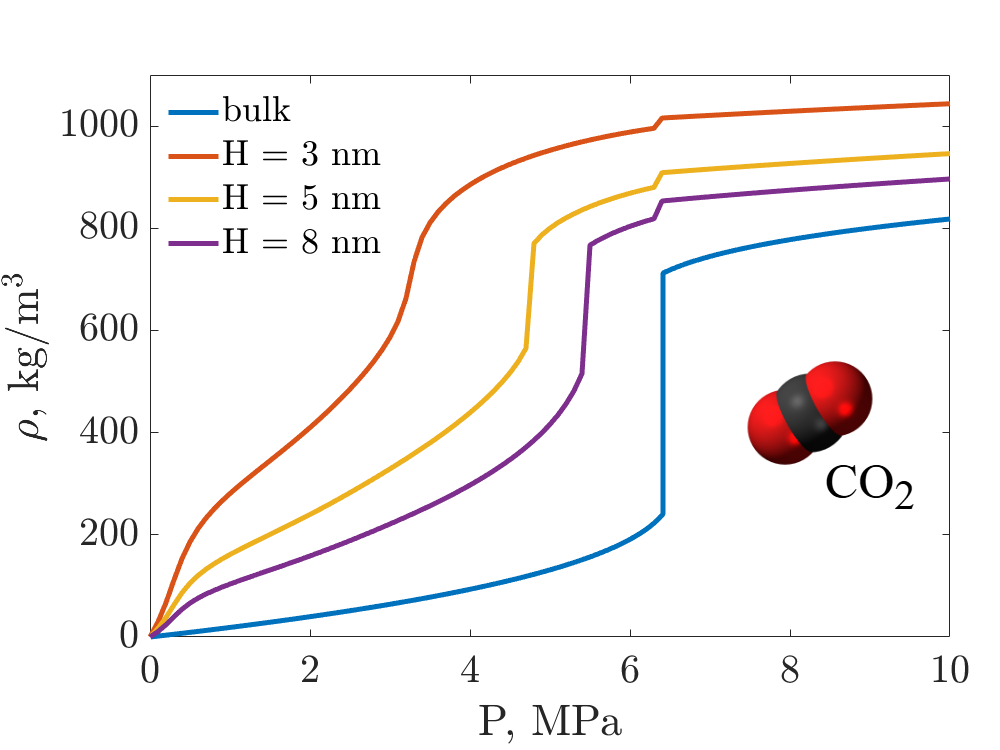}
    \caption{PVT for $CO_2$ at T = 298~K in the pores H = 3, 5, and 8~nm during pressure increase in the bulk.}
    \label{fig:comparison_rho_pore_bulk}
\end{figure}

To begin with, we consider a case study of permeability drop for carbon dioxide at T = 298~K inside the 'uni-500' sample. As illustrated in Fig. \ref{fig:comparison_rho_pore_bulk}, the PVT characteristics of carbon dioxide at T = 298~K in the pores H = 3, 5, and 8~nm deviate from the bulk PVT behavior. As one can see, PVT depends on the pore size, and wider pores provide fluid PVT properties closer to its bulk behavior. The density-pressure curves have a step-like pattern that reflects the phase transition in the pore, i.e., capillary condensation at pressure increase. Additionally, at the pressure of phase transition in the bulk P$_{ph.tr.}^{bulk}$ = 6.4121~MPa for T = 298~K, we see a small increase in average density in the pores. It is explained by the connection of pores with the bulk, so when the fluid in the bulk undergoes a phase transition, it influences the fluid state in the pore. The results show that capillary condensation in the pores occurs at lower pressure values than the pressure of phase transition in the bulk, and it becomes more significant for smaller pores. 

\begin{figure}
    \centering
    \includegraphics[width=1\linewidth]{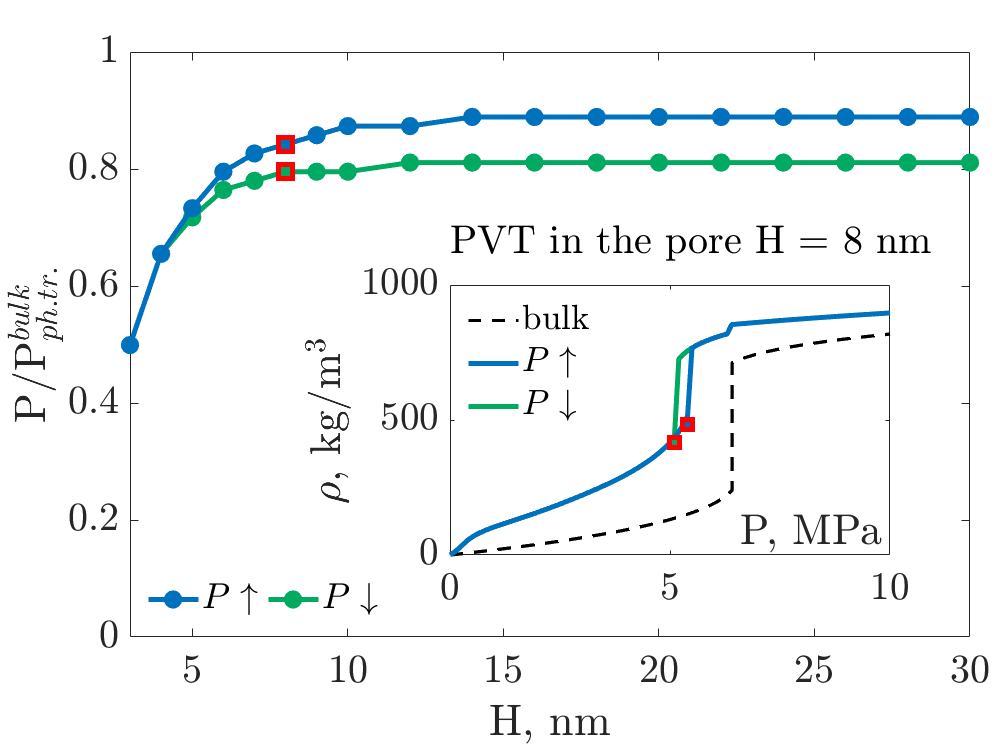}
    \caption{Pressures of capillary condensation (blue) and evaporation (green) for $CO_2$ at T = 298~K in the pores H = 3--30~nm. Inset: capillary hysteresis at H = 8~nm.}
    \label{fig:pvt_hyst}
\end{figure}

\begin{figure*}
    \centering
    \includegraphics[width=1\linewidth]{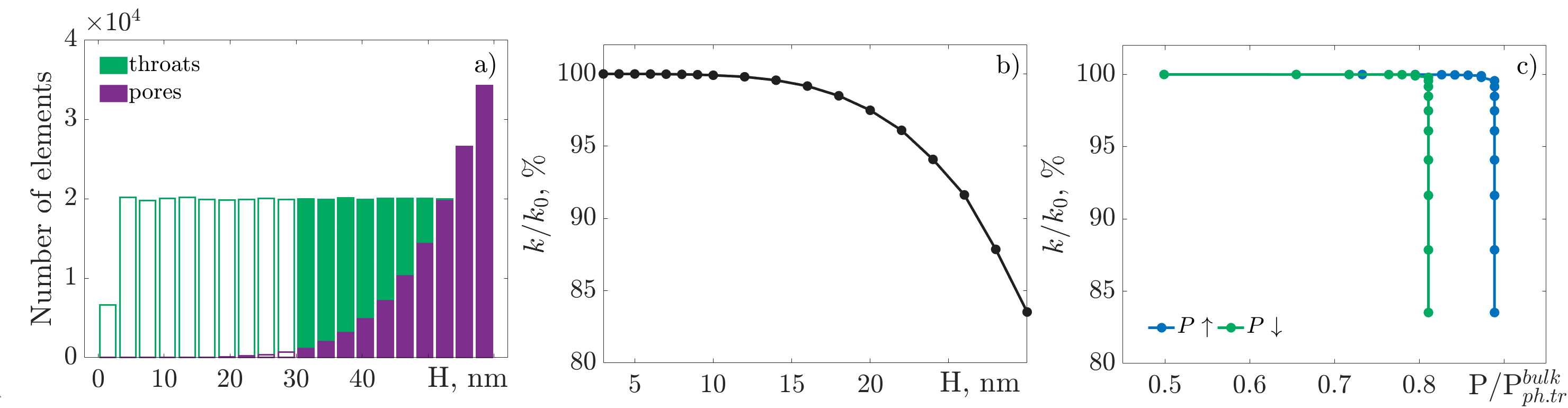}
    \caption{a) PSD and TSD of 'uni-500' sample, where white bars notes blocked pores and throats up to 30~nm and the colored bars notes the pores and throats accessible for the gas flow, b) Permeability drop depending on the width of blocked pores for $CO_2$ filtration in the sample 'uni-500' at T = 298~K, c) Permeability drop depending on the pressures of capillary condensation and evaporation for $CO_2$ filtration in the sample 'uni-500' at T = 298~K.}
    \label{fig:psd_lin}
\end{figure*}

After that, we collect the results of the pressures for capillary condensation during pressure increase and capillary evaporation during pressure decrease inside pores with widths up to 30 nm to investigate the capillary hysteresis phenomenon. These results for carbon dioxide at T = 298~K and pores in the range H = 3--30~nm are shown in Fig. \ref{fig:pvt_hyst}. The inset illustrates the PVT of carbon dioxide in the pore H = 8~nm during pressure increase and decrease, reflecting capillary condensation and evaporation. To represent the gap of capillary hysteresis, we considered the beginning of phase transition during capillary condensation and the end of phase transition during capillary evaporation, highlighted as red squares on Fig. \ref{fig:pvt_hyst}. Notably, in the smaller pores with widths of 3 and 4~nm, no capillary hysteresis is observed. Capillary hysteresis occurs in the pores bigger than 5~nm and grows with increasing pore width (slight narrowing in the area of 12~nm is related to the coarse computational mesh with a pressure step of 0.1 MPa). The pressures at which capillary condensation and evaporation occur also demonstrate an upward trend with increasing pore width. This behavior aligns with previous experimental observations and theoretical predictions obtained by cDFT calculations \cite{ravikovitch2001characterization, neimark2003bridging, monson2012understanding}. A detailed discussion regarding the nature of capillary hysteresis is provided in the Discussion \ref{sec:disc_A} section.

To reflect the impact of capillary hysteresis during PNM simulations, we block the pores and throats where the condensate phase is formed at a particular pressure. As a result of PNM simulations, we obtain a relationship between the initial single-phase permeability $k_0$ of a sample and its permeability in the presence of capillary condensation $k$, shown in Figure \ref{fig:psd_lin}. We consider a randomly generated artificial sample labeled 'uni-500' characterized by a uniform distribution of throat sizes, with pore and throat size distributions shown in Figure \ref{fig:psd_lin}a. The white columns in Figure \ref{fig:psd_lin}a indicate the pores and throats that are considered to be blocked for fluid flow. The range of pores we consider to be blocked, from 3 to 30~nm, constitutes mostly throats and a small number of pores. Figure \ref{fig:psd_lin}b shows permeability reduction corresponding to the width of the blocked pores. This result shows the contribution of each pore size to the overall flow within the sample. We observe that the permeability drop could achieve about 17\% if pore sizes up to 30~nm are blocked by a condensate phase. The obtained permeability drop is mostly attributed to a blockage of pores bigger than 10~nm. The deviation of permeability while blocking pores smaller than 10~nm is lower than 1\% in this sample. Furthermore, due to capillary hysteresis, where phase transition in the pore occurs at different pressure values during pressure increase and decrease, we observe hysteresis in the relationship between permeability and pressure, as shown in Figure \ref{fig:psd_lin}c. This finding indicates that the permeability of the sample is influenced by the thermodynamic processes occurring within the system if the condensate phase is formed and capillary forces dominate the viscous one.

\subsection{Temperature impact on permeability hysteresis} \label{sec:perm_T}

\begin{figure*}
    \centering
    \includegraphics[width=1\linewidth]{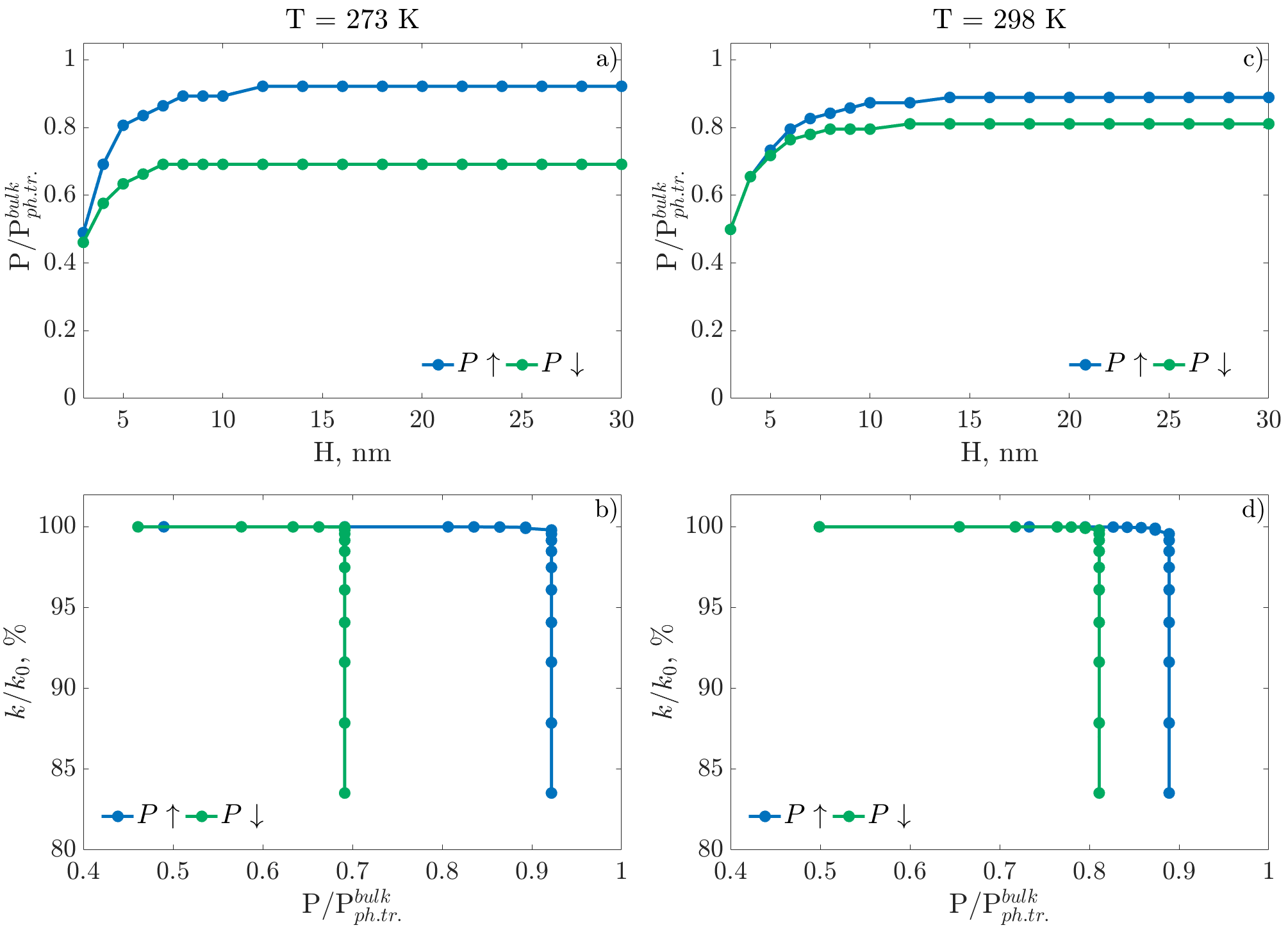}
    \caption{a) Pressures of capillary hysteresis for $CO_2$ depending on pore width at T = 273~K, b) Permeability hysteresis for $CO_2$ gas flow in the sample 'uni-500' depending on pressure at T = 273~K, c) Pressure of capillary hysteresis depending on pore width for $CO_2$ at T = 298~K, d) Permeability hysteresis for $CO_2$ gas flow in the sample 'uni-500' depending on pressure at T = 298~K.}
    \label{fig:condensation_hyst}
\end{figure*}

In this subsection, we examine the temperature impact on permeability hysteresis. It is well known that capillary hysteresis within the pores is temperature-dependent \cite{neimark2000adsorption, li2014phase, barsotti2016review, barsotti2020capillary}. The capillary hysteresis becomes more pronounced for lower temperatures. For the existence of capillary condensation in the pores, we need to consider a system temperature lower than the critical temperature of the fluid in the bulk $T<T_c$. Generally, the critical temperature for fluid inside a pore is lower than its critical temperature in the bulk. It is also known that the temperature for the existence of capillary hysteresis is lower than the critical temperature in the pore \cite{neimark2003bridging, barsotti2016review}. However, all these parameters are also dependent on the pore width \cite{balbuena1993theoretical, neimark2000adsorption, monson2012understanding, neimark2003bridging}. If we incorporate cDFT calculation results for fluid capillary hysteresis into the PNM calculations, considering pore blockage by a condensate, then the permeability hysteresis can be obtained and investigated at different temperatures. 

Figure \ref{fig:condensation_hyst} presents the results on capillary hysteresis and corresponding permeability hysteresis for carbon dioxide at temperatures T = 273~K and T = 298~K. The critical temperature for carbon dioxide is T$_c$ = 304.1282~K. During our calculations, we use a randomly generated 'uni-500' sample. For both series of calculations, we considered carbon dioxide PVT behavior in the pore size range of 3--30~nm. The results indicate a direct correlation between the pressures of capillary hysteresis and the corresponding pressures of permeability hysteresis. Similar to capillary hysteresis behavior with temperature, lower temperatures enhance the permeability hysteresis. It means that at $T<T_c$, thermodynamic processes would impact the permeability of the sample. Our findings reveal that temperature influences only the pressure at which a permeability drop occurs, but it does not change the magnitude of the permeability drop. The value of the permeability drop is probably controlled only by sample geometry.

\subsection{Impact of element size distributions on permeability hysteresis} \label{sec:perm_psd}

\begin{figure*}
    \centering
    \includegraphics[width=1\linewidth]{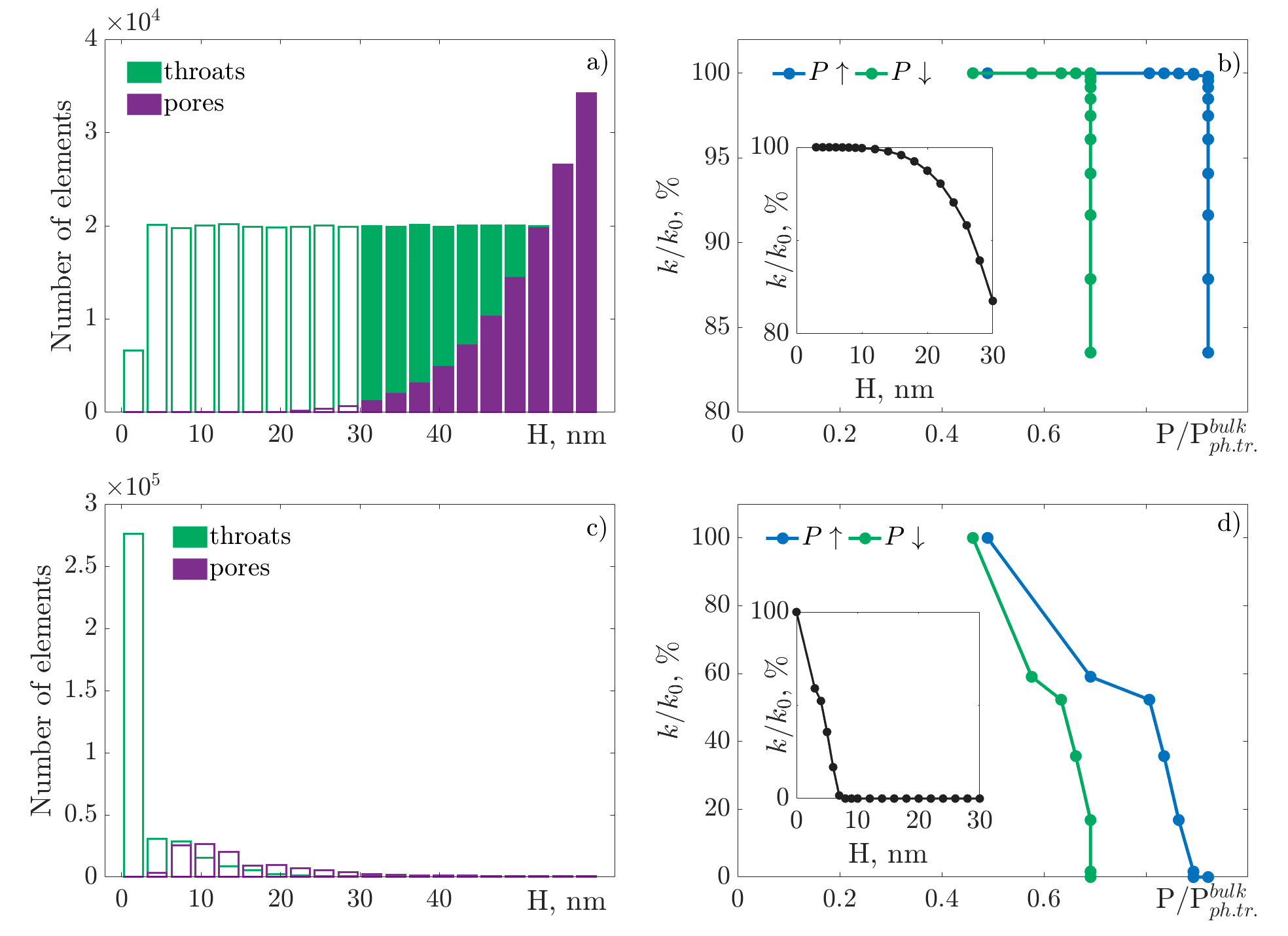}
    \caption{PSD and TSD for the samples a) 'uni-500', c) 'vm-500' alongside, where white bars notes blocked pores and throats up to 30~nm and the colored bars notes the pores and throats accessible for the gas flow, alongside with the results of permeability drop and hysteresis caused by $CO_2$ capillary hysteresis at T = 273~K in the samples b) 'uni-500', d) 'vm-500'.}
    \label{fig:k_drop_samples}
\end{figure*}

Numerous studies have demonstrated that porous structures with different PSDs and TSDs exhibit significantly varying filtration and storage properties  \cite{panda1994estimation, Mehmani2014, Prodanovi2014, mehmani2014application, xu2022pore}. 
To investigate how the permeability is controlled by sample geometry, we perform PNM calculations of permeability accounting for capillary hysteresis in two artificial samples with different TSDs. The effect of TSD on permeability reduction and hysteresis was evaluated considering carbon dioxide capillary condensation and evaporation at temperature T = 273~K within pores up to H = 30~nm. We consider two randomly generated samples: one with a uniform TSD noted as 'uni-500', and another with a TSD and PSD taken from the Vaka Muerta formation sample \cite{berthonneau2021nanoscale}, referred to as 'vm-500'. The samples PSDs, TSDs, and results of the permeability drop and hysteresis are illustrated in Figure \ref{fig:k_drop_samples}. Our findings indicate notable differences in permeability reduction and hysteresis between the two samples considered. For the 'uni-500' sample, capillary condensation blocked approximately 1/2 of the throats and a small number of pores, resulting in a permeability decrease of approximately 17\%. In contrast, for the 'vm-500' sample, capillary condensation blocked the majority of pores and throats, leading to a complete stop of fluid flow and a permeability drop to zero when the pores and throats with sizes less than 10~nm are blocked. Thus, the results proved that pore space geometry is a critical parameter to determine the permeability when considering capillary condensation effects within nanopores.

\subsection{Impact of sample linear size on permeability drop} \label{sec:perm_size}

\begin{figure}
    \centering
    \includegraphics[width=1\linewidth]{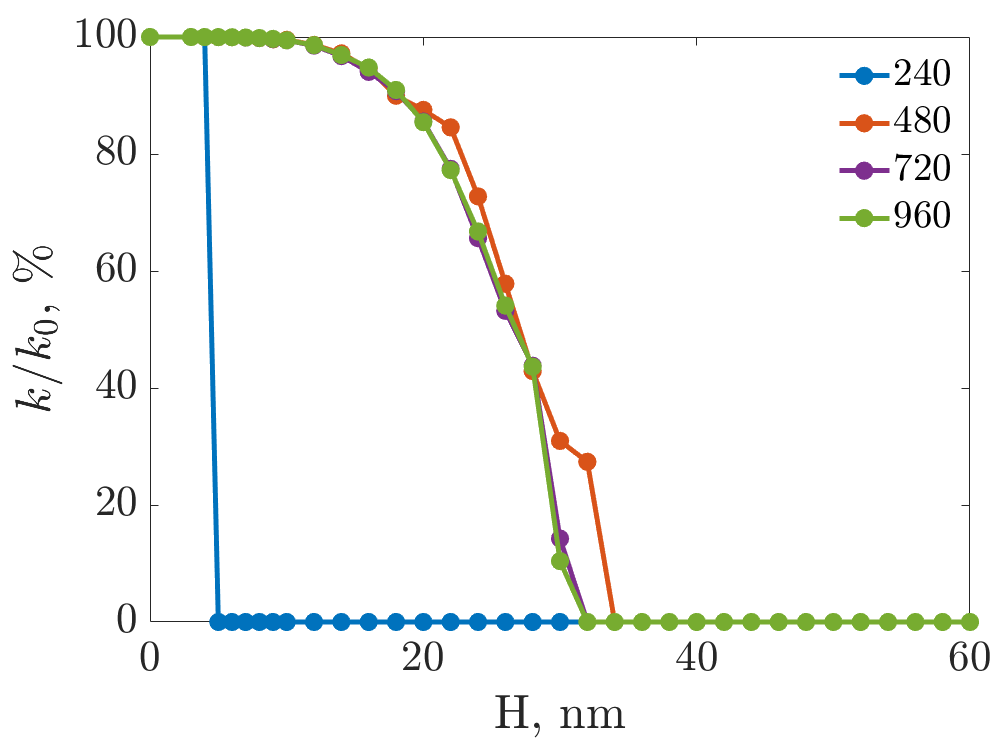}
    \caption{Permeability drop for 'uni' samples with various sizes by capillary condensation of $CO_2$ at T = 273~K.}
    \label{fig:sizes}
\end{figure}

For simulation of pore-scale filtration, it is essential to ensure that the results accurately reflect the behavior of a larger system, addressing the concept of representative elementary volume (REV). To evaluate the required REV size, we examined randomly generated 'uni' samples of varying sizes: 240, 480, 720, and 960. It reflects the physical size of the considered sample and, consequently, the number of pores and throats in the PNM, given in Table \ref{tab:sizes}. Figure \ref{fig:sizes} illustrates the permeability drop observed in the 'uni' samples with different sizes, accounting for capillary condensation of carbon dioxide at temperature T = 273~K. Notably, the smallest pore structure sample, 'uni-240', exhibited a significant deviation in permeability drop compared to the larger samples.  This pore network is definitely smaller than the size required for REV and is unsuitable for filtration simulation. Such deviation is related to the fact that when the particular pores are blocked, it can block the whole way for fluid flow, and for a smaller sample, flow disruption occurs earlier than in larger systems. Our analysis suggests that a minimum size of approximately 480 is necessary for representativity, although this threshold may also depend on additional characteristics of the pore-scale geometry and fluid properties.

\renewcommand{\arraystretch}{1.1} %% increase table row spacing
\renewcommand{\tabcolsep}{0.2cm} %% increase table column spacing
\begin{table}[h!]
\caption{Sample size characteristics with uniform TSD.}
\label{tab:sizes}
\begin{ruledtabular}
\begin{tabular}{lcc}

\multirow{2}{*}{Linear size} &
  \multirow{2}{*}{Number of pores} &
  \multirow{2}{*}{Number of throats} \\
            &                &                \\
                          \midrule
\multicolumn{1}{c}{240}  & \multicolumn{1}{c}{512}          & 1472\\ 
\multicolumn{1}{c}{480}  & \multicolumn{1}{c}{4096}          & 12032\\ 
\multicolumn{1}{c}{720}  & \multicolumn{1}{c}{13824}          & 40896\\ 
\multicolumn{1}{c}{960}  & \multicolumn{1}{c}{32768}          & 97280\\ 
\end{tabular}
\end{ruledtabular}
\end{table}

\begin{figure}
    \centering
    \includegraphics[width=1\linewidth]{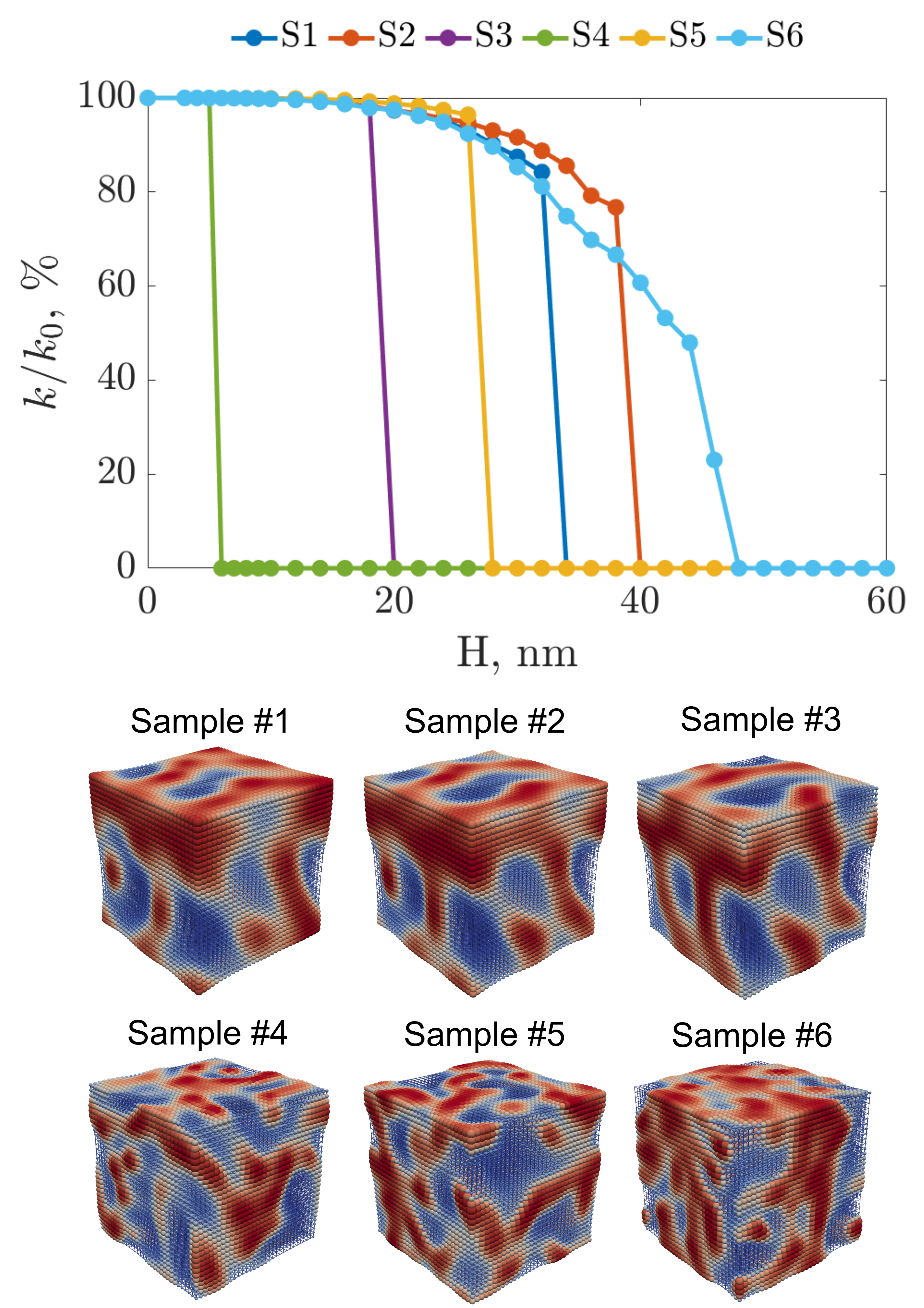}
    \caption{Permeability drop caused by capillary condensation of $CO_2$ at T = 273~K for 6 different samples artificially generated from uniform distribution with space-correlated pore arrangement by a Gaussian field.}
    \label{fig:structure_result}
\end{figure}

\subsection{Impact of sample structural correlations on permeability drop} \label{sec:perm_struct}

Now, we investigate the impact of sample structural correlations to ensure that the knowledge of sample TSD and PSD is sufficient for the simulation of fluid transport through multiscale porous media, accounting for nanoconfined phase behavior. To examine this, we analyzed 6 different samples randomly generated using uniform TSD, where we vary the pore space geometry using Gaussian field correlation. The parameters for a Gaussian field used for the patterns of porous space are given in Table \ref{tab:GaussFielParams}. Figure \ref{fig:structure_result} illustrates the constructed pore space geometry of the considered samples and the corresponding permeability drop caused by capillary condensation of carbon dioxide at T = 273~K up to 60~nm. As can be seen from Fig. \ref{fig:structure_result}, the geometry of porous space is varied by different rearrangements of large and small pores and throats, resulting in different characters of porous space heterogeneity, keeping pore connectivity constant.
Particularly, variations in the location and size of zones with large and small porosity can be treated as equivalent to different topologies of porous space. Notably, although all samples have similar TSD and PSD, the behavior of permeability reduction varies significantly. The permeability of each sample turns to 0 at different values of pore size that are blocked by a condensate phase. It is related to that since the pores are rearranged differently, when the pores with a particular pore size are blocked by a condensate, it leads to a different topology of the resulting porous space available for the gas flow. This observation reveals that TSD and PSD do not provide enough information about the porous space structure required to accurately simulate multiscale flow filtration. Even if the samples have similar element size distributions, variations in their spatial arrangements can lead to different flow behaviors. A similar conclusion about the role of porous space topology was made in the work \cite{mehmani2014application}.

\renewcommand{\arraystretch}{1.1} %% increase table row spacing
\renewcommand{\tabcolsep}{0.2cm} %% increase table column spacing
\begin{table} [!h]
\caption{Gaussian field parameters.}
\label{tab:GaussFielParams}
\begin{ruledtabular}
\begin{tabular}{cccc}

\multirow{2}{*}{Sample №} &
  \multirow{2}{*}{$P_0$} &
  \multirow{2}{*}{$\zeta$} &
  \multirow{2}{*}{$k_1$}\\
            &                &                \\
                          \midrule
\multicolumn{1}{c}{1}  & \multicolumn{1}{c}{1.0}          & 0 & 1\\ 
\multicolumn{1}{c}{2}  & \multicolumn{1}{c}{1.0}          & 0.2 & 1\\ 
\multicolumn{1}{c}{3}  & \multicolumn{1}{c}{1.0}          & 0.4 & 1\\ 
\multicolumn{1}{c}{4}  & \multicolumn{1}{c}{1.0}          & 0 & 2\\ 
\multicolumn{1}{c}{5}  & \multicolumn{1}{c}{1.0}          & 1 & 2\\ 
\multicolumn{1}{c}{6}  & \multicolumn{1}{c}{1.0}          & 0 & 3\\ 
\end{tabular}
\end{ruledtabular}
\end{table}

\section{Discussion}

\subsection{Capillary hysteresis} \label{sec:disc_A}

\begin{figure}
    \centering
    \includegraphics[width=1\linewidth]{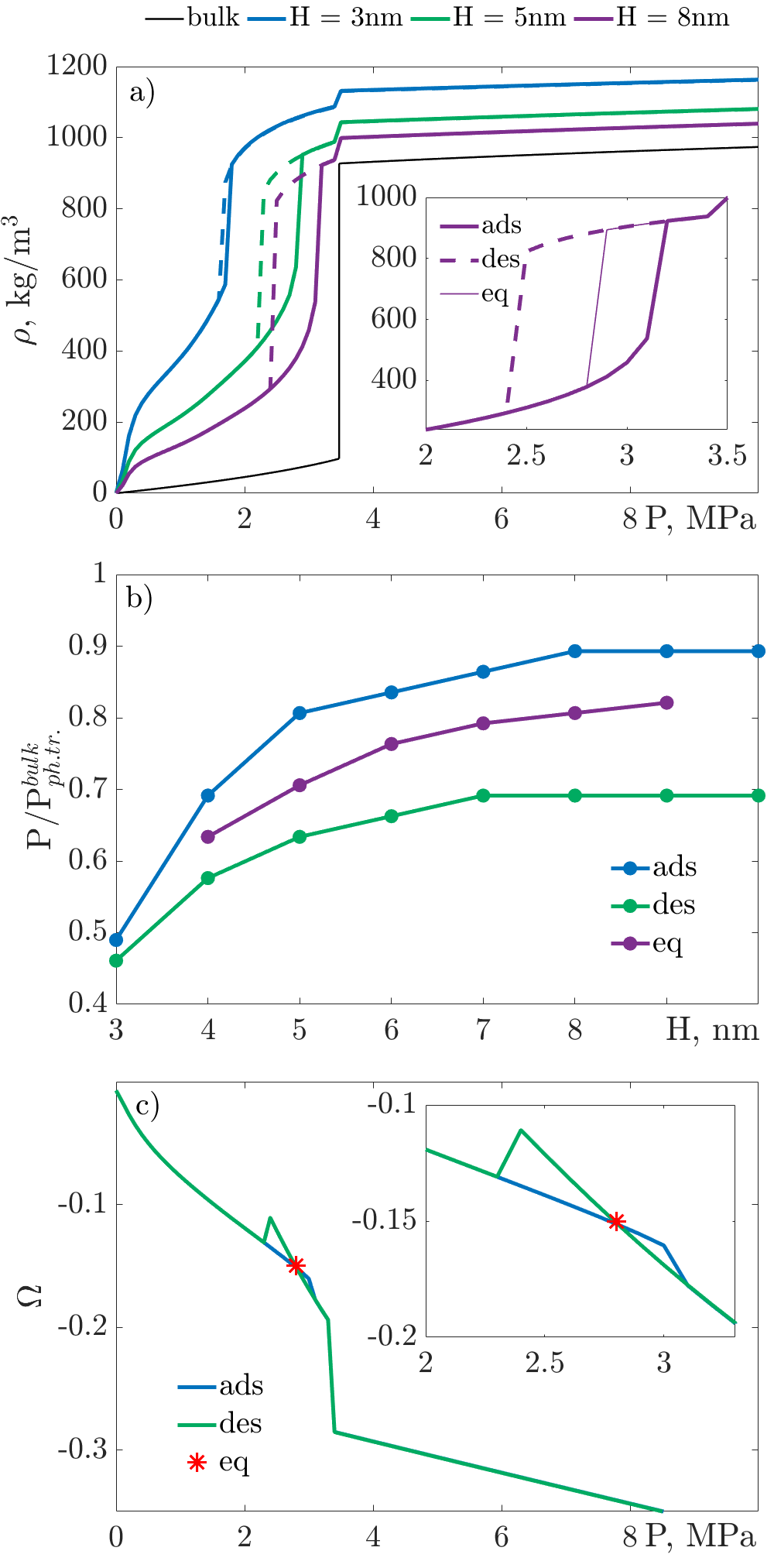}
    \caption{At the top: pore PVT for carbon dioxide at T = 273~K and in the inset non-equilibrium adsorption and desorption along with equilibrium adsorption in the pore H = 8~nm; At the center: Pressure of capillary condensation during adsorption, desorption and equilibrium adsorption compared with bulk value of pressure for phase transition at T = 273~K; At the bottom: free energy loop for carbon dioxide inside the slit-like pore with H = 8~nm at T = 273~K.}
    \label{fig:ads_des_hyst}
\end{figure}

To investigate capillary hysteresis in the nanopores, we perform cDFT calculations of fluid pore PVT during pressure increase and decrease. These curves show at which thermodynamic conditions capillary condensation and evaporation inside the pores occur, respectively. Figure \ref{fig:ads_des_hyst}a illustrates the PVT curves for carbon dioxide at T = 273~K in the pores H = 3, 5, and 8~nm during pressure increase and decrease, alongside the bulk PVT curve. Our results indicate that in smaller pores, the capillary hysteresis loop shifts from the bulk pressure of phase transition to lower pressures, and the gap between pressures of capillary condensation and evaporation reduces. Important to note that the pore PVT curves during pressure increase and decrease are similar to non-equilibrium adsorption-desorption isotherms up to the normalization factor of fluid density. If so, similar behavior was also previously observed experimentally and by using cDFT calculations \cite{neimark2003bridging, ravikovitch2001characterization, monson2012understanding, li2014phase}. Remarkably, such behavior of a hysteresis loop is common for H1 type of hysteresis resulting from the formation of metastable states of fluid \cite{barsotti2020capillary}.

Previous studies usually focus on fluid phase behavior in smaller pore sizes, but what will happen if we consider wider pores till it turns to bulk? The capillary hysteresis is the result of non-equilibrium processes happening only in the micropores \cite{balbuena1993theoretical, neimark2000adsorption, neimark2001capillary, sangwichien2002density}. While inside macropores and at the bulk conditions, fluid exhibits equilibrium behavior without hysteresis. According to work \cite{jagiello20132d}, the adsorption curve approaches a bulk one with pore width increase. We obtain similar results; however, the desorption curve does not exhibit a similar trend toward the bulk value of the phase transition. Particularly, it grows with pore width increase, but it approaches pressure values much smaller than the pressure of phase transition at the bulk (see Fig. \ref{fig:ads_des_hyst}b). How then does capillary hysteresis transform into the bulk behavior? Actually, the capillary hysteresis appears only in the so-called "mesopores" according to IUPAC classification, which are about 2-50~nm. While these values should be treated as reference values, it is important to acknowledge that the specific values can vary depending on the fluid and pore wall characteristics. For example, in the results presented in the subsection Result \ref{sec:perm_drop}, we obtained that the minimum pore size for the existence of capillary hysteresis for carbon dioxide at T = 298~K is around 5 nm for the carbon-wall pore. However, our approach doesn't allow us to estimate the maximum value of pore width for the existence of capillary hysteresis. These values can probably be taken only from the experimental data.

To improve the analysis, in addition to the nonequilibrium curves, we calculated the pressures for an analogous to an equilibrium adsorption curve that lies within the hysteresis loop (see the inset of Fig. \ref{fig:ads_des_hyst}a). To achieve the equilibrium curve, we analyzed the free energy values and identified the global minimum for the fluid confined within the pore. In Fig. \ref{fig:ads_des_hyst}c, we show the free energy loop obtained during nonequilibrium adsorption and desorption processes. The moment when these curves intersect corresponds to the moment of phase transition during equilibrium adsorption inside pores \cite{neimark2003bridging, monson2012understanding, neimark2001capillary}. In Fig. \ref{fig:ads_des_hyst}b, we present the pressure values for fluid phase transition, i.e., capillary condensation and capillary evaporation, during non-equilibrium adsorption and desorption, and equilibrium one inside nanopores with different widths. Equilibrium values of pressures for phase transition occur between the pressure values of fluid phase transition during nonequilibrium adsorption and desorption \cite{neimark2003bridging}. Consequently, we can assume that as the pore width increases, the pressure at which capillary condensation occurs during equilibrium adsorption approaches the bulk value, similar to non-equilibrium adsorption behavior.

Predicting the specific pressure at which capillary condensation occurs within a real porous structure poses significant challenges. Previous studies \cite{ravikovitch2001characterization,neimark2000adsorption, neimark2001capillary, thommes2015physisorption, haidar2024small} indicated that in some cases experimental adsorption and desorption results align with theoretical non-equilibrium curves, but sometimes experimental curves converge with the theoretical equilibrium curve. However, it is commonly treated that adsorption is metastable and desorption is in equilibrium \cite{haidar2024small}. This variability can be attributed to specific factors influencing the system's non-equilibrium behavior, such as the local structural characteristics of pore walls, which are usually unknown \cite{haidar2024small}. Additionally, local chemical and geometrical heterogeneities of the pore walls or the presence of various impurities in the fluid can significantly affect condensation behavior \cite{khlyupin2017random, aslyamov2017density,aslyamov2019theoretical, guo2026co2}. Moreover, when considering the whole porous structure, the following propagation of condensate phase formation depends on the connectivity of the porous space, and this factor can also influence fluid phase behavior in the porous space.

\subsection{Future perspective} \label{sec:disc_B}

The present study deliberately isolates one specific nanoscale mechanism and follows its impact up to the scale of the entire multiscale pore system: the effect of confinement-induced phase behavior of a single-component fluid on the effective permeability of a heterogeneous pore network. In particular, the model focuses on capillary condensation and evaporation in nanopores, including the associated hysteresis, and examines how the resulting pore blockage modifies the conducting topology of the network. This simplified formulation is intentional. Transport in organic-rich multiscale porous media is governed by many nanoscale effects, but separating the role of capillary condensation hysteresis is necessary in order to quantify its standalone contribution to permeability reduction and to the emergence of permeability hysteresis at the sample scale. Accordingly, the present framework should be interpreted not as a complete description of nanoscale flow in shale-like materials, but as a controlled analysis of one physically important mechanism that is often discussed in studies of confined thermodynamics and phase behavior, yet is still rarely integrated explicitly into a network-scale filtration model.

At the same time, the proposed framework can be extended in a natural way toward a more complete transport description \cite{dinariev2017modeling, stierle2021hydrodynamic, balashov2025two}. In its current form, the model does not include several well-established nanoscale flow effects that may substantially modify gas transport in the porous media. These effects include slip flow at solid boundaries, Knudsen diffusion, surface diffusion of the adsorbed phase, wall-material-dependent accommodation, and confinement-induced changes in local density, viscosity, and velocity profiles \cite{javadpour2021gas, wu2015model, yousefi2023exploring, xiao2024adsorption, ma2014pore, zhang2015micro}. Prior studies show that gas transport in shale nanopores may deviate strongly from classical no-slip Poiseuille flow because the relative contribution of viscous flow, slip transport, Knudsen transport, and adsorption-controlled mobility depends on pore size, pressure, wall chemistry, and fluid-rock interactions \cite{javadpour2021gas, ma2014pore, zhang2015micro, nan2020slip}. They also indicate that confined fluids can exhibit near-wall densification and altered rheological behavior, which in turn modify the effective flow profile \cite{wang2017multiscale, wang2020multiscale, yu2020multiscale}. In future developments, these effects could be incorporated as complementary mechanisms controlling the conductance of pores and throats that remain hydraulically connected. In such an extended formulation, capillary condensation would determine which network elements are removed from gas transport, whereas slip, Knudsen, surface-diffusion, and confinement-corrected rheology would determine how the residual connected subnetwork transmits flow.

A particularly promising direction is the extension from single-component systems to multicomponent confined fluids. In real shale and other organic-rich porous media, phase behavior becomes substantially more complex once fluid composition is considered. Competitive adsorption, selective enrichment of heavier components, redistribution of species between organic nanopores and larger inorganic pores, and composition-dependent phase transitions may all arise simultaneously \cite{sobecki2019phase, wang2019competitive, wang2023pore, chen2024phase, chen2021pore, bi2019molecular}. Under these conditions, nanopores do not merely act as passive regions with modified local PVT behavior; they may actively fractionate the fluid mixture and thereby change the composition, density, viscosity, and phase envelope of the mobile fluid in larger pores and fractures \cite{sobecki2019phase, wang2023pore, chen2024phase, wan2024phase, chen2021pore, bi2019molecular}. As a result, even the transport properties of apparently macroporous domains may be indirectly controlled by nanoconfined thermodynamics. In this sense, the mechanism analyzed in the present work can be viewed as the single-component limit of a broader class of multiscale compositional effects, in which local thermodynamic partitioning between pore subsystems governs the global filtration process. Recent studies strongly support the importance of these multicomponent effects, especially for systems involving methane, carbon dioxide, water, and heavier hydrocarbons \cite{sobecki2019phase, wang2019competitive, wang2023pore, chen2024phase, wan2024phase, chen2021pore, bi2019molecular}.

To describe such phenomena efficiently and with sufficient physical validity, further progress will likely require hybrid molecular-to-mesoscale workflows combining molecular dynamics and classical density functional theory. Pure molecular dynamics remains indispensable for resolving local transport mechanisms, wall-specific mobility, and material-dependent interfacial effects, but it becomes prohibitively expensive when one needs to explore a broad range of pore sizes, thermodynamic conditions, and mixture compositions, especially in open systems that must remain in equilibrium with larger pores \cite{xiao2024adsorption, vaganova2022linking}. By contrast, cDFT provides a computationally efficient route to equilibrium density profiles, adsorption, layering, and phase transitions in confinement \cite{vaganova2022linking, de2024classical}. A practical strategy, therefore, is to use cDFT to map equilibrium thermodynamics across parameter space, while targeted MD simulations are used to calibrate transport behavior, validate near-wall structure and mobility, and learn accurate PVT behavior for confined mixtures \cite{vaganova2022linking, de2024classical}. This division of roles is especially attractive for multicomponent systems, where direct MD sampling over all relevant pore sizes and compositions would be prohibitively costly \cite{vaganova2022linking}. In this sense, cDFT models trained on a limited but representative set of MD simulations may become a key ingredient of next-generation multiscale models for confined compositional transport \cite{vaganova2022linking, de2024classical}.

Finally, the development of physically grounded multiscale transport models must proceed along with the development of reliable methods for three-dimensional pore-space reconstruction and structural uncertainty quantification \cite{cherkasov2024towards, Kulygin2024, cherkasov2021adaptive}. One of the central conclusions of the present work is that pore and throat size distributions alone are insufficient to predict the filtration response under confinement-induced pore blocking; the spatial arrangement, connectivity, and topology of the pore space also play a decisive role. However, current experimental techniques rarely provide complete and unique 3D information at the relevant scales. The limitations of mercury porosimetry, adsorption-based techniques, CT, SEM, FIB-SEM, and electron tomography are well known, as well as the persistent problem of representativity and the difficulty of identifying higher-order structural organization from limited digital samples. This implies that future studies should move beyond single deterministic digital structures toward stochastic or ensemble-based reconstructions consistent with the available measurements \cite{xiao2024three, lee2024development, fu2025computational}. Such ensembles would allow one to quantify the uncertainty of predicted permeability, hysteresis width, and critical blocking thresholds with respect to unresolved geometric variability \cite{xiao2024three, fu2025computational}. For multiscale porous materials, uncertainty in structure is therefore not merely a technical drawback but an intrinsic part of the modeling problem itself.
\section{Conclusion}

We present a novel upscaling framework for estimating the permeability of multiscale porous systems, incorporating the effects of nanoconfined fluid phase behavior. The framework integrates fluid PVT in the nanopore, obtained by classical Density Functional Theory calculations, into quasi-static Pore Network Modelling. The confinement effect leads to capillary condensation of a fluid in the pore at lower pressures than in the bulk. When the fluid in the pore is in the liquid phase, the pore is considered to be blocked and does not account for the filtration simulation within the PNM algorithm. This study results in the following observations:
\begin{enumerate}
    \item The consideration of nanoconfined fluid phase behavior during simulation of gas filtration in the multiscale porous media leads to permeability decrease under the assumption of low capillary number.
    \item Capillary hysteresis inside nanopores leads to similar hysteresis of sample permeability.
    \item The size of the porous sample must be large enough to produce representative results regarding fluid filtration and the influence of nanoscale effects.
    \item The structural characteristics of the sample significantly influence permeability reduction caused by nanoconfined phase behavior; even samples with similar pore and throat size distributions show different filtration results due to variations in of structural correlation that reflect  multiscale spatial heterogeneity.
    
\end{enumerate}
\section{Acknowledgements}

The authors thank the Russian Science 
Foundation, Russia (Grant No. 25-13-00313) for financial support. 

\newpage
\bibliography{references}
\appendix

\section{Classical Density Functional Theory details}\label{sec:Appendix_DFT}

In this section, we provide exact formulations for the Helmholtz free energy functional $F \left[ \rho\right]$, the external potential $V^{ext}$, the chemical potential $\mu$, and cDFT EoS, i.e., the relation between pressure and density, used during cDFT calculations. In the expression for the Helmholtz free energy functional $F \left[ \rho\right]$, to treat molecular repulsion, which is described by hard-sphere interactions $F^{hs} \left[ \rho\right]$, we use the Fundamental Measure Theory approach (FMT) \cite{rosenfeld1989free}. The interactions of molecular attraction $F^{att} \left[ \rho\right]$ are formulated by the Mean Field Approximation framework (MFA), as in the work \cite{ravikovitch2001density}. Then, the Helmholtz free energy functional $F \left[ \rho\right]$ is given by following equations: 
\begin{align}
    &F =  F^{id} + F^{hs} + F^{att} \label{eq:sum}\\
    &F^{id} = k_B T\int d\vec{r}\,\rho\left(\vec{r}\right)\left(\ln{{(\Lambda}^3 \rho\left(\vec{r}\right))}-1\right)\label{eq:id}\\
    &F^{hs} = k_B T\int d\vec{r}\,\Phi\left[n_\alpha\left(\rho\left(\vec{r}\right)\right)\right]\label{eq:HS}\\
    &F^{att}=k_B T \iint d\vec{r}\,\rho\left(\vec{r}\right)d\vec{r}^\prime\rho\left(\vec{r}^\prime\right)U^{att}(\vert\vec{r}-\vec{r}^\prime\vert) \label{eq:att}
\end{align}
where $k_B$ is the Boltzmann constant, $T$ is the system temperature, $\Lambda = \frc{h}{\sqrt{2\pi mT}}$ the thermal de Broglie wavelength, with $h$ the Planck constant and $m$ the mass of the molecule, $\Phi\left[n_\alpha\left(\rho\left(\vec{r}\right)\right)\right]$ is the Rosenfeld functional and $n_\alpha$ (scalar $\alpha$ = 0,1,2,3; vector $\alpha$ = 1,2) the weighted density, given by:

\begin{eqnarray}\label{eq:rosienfield}
    \Phi =  -n_0 \ln{\left(1-n_3\right)} & + & \frac{n_1 n_2- \vec{n_1}\cdot\vec{n_2}}{1-n_3} \nonumber\\  & & +  \frac{n_2^3-3n_2 \vec{n_2}\cdot\vec{n_2}}{24\pi\left(1-n_3\right)^2},    
\end{eqnarray}

\begin{equation}\label{eq:weighted_dens}
    n_\alpha \left(\vec{r}\right)= \int d^3r^\prime \rho\left(\vec{r}^\prime\right)\omega_\alpha\left(\vec{r}-\vec{r}^\prime\right),
\end{equation}
where $\omega_\alpha$ are the weight functions; $\omega_3\left(\vec{r}\right)=\theta\left(R-r\right)$, $\omega_2\left(\vec{r}\right)=\delta\left(R-r\right)$, ${\vec{\omega}}_2\left(\vec{r}\right)=\frac{\vec{r}}{r}\delta\left(R-r\right)$, $\omega_1= \frac{\omega_2}{4\pi R}$, $\omega_0 = \frac{\omega_2}{4\pi R^2}$, ${\vec{\omega}}_1= \frac{{\vec{\omega}}_2}{4\pi R}$, $\delta$ and $\theta$ are the Dirac delta function and the Heaviside step function, respectively, $R$ is particle radius.
The potential of intermolecular interactions $U^{att}$ is expressed as:
\begin{equation}
    U^{att}\left(r\right)=\ \left\{
    \begin{matrix}
        -\epsilon&r<\lambda\\
        U^{LJ}&\lambda<r<r_{cut}\\
        0&r>r_{cut}\\
    \end{matrix}\right.
\end{equation}
\begin{equation}
    U^{LJ}=4\epsilon\left(\left(\frac{\sigma}{r}\right)^{12}- \left(\frac{\sigma}{r}\right)^6\right).
\end{equation}
with $r = \vert\vec{r}-\vec{r}^\prime\vert$, $\epsilon$ and $\sigma$ the effective intermolecular interaction parameters that are calculated by fitting cDFT EoS on isothermal data from NIST Chemistry WebBook, $\lambda = 2^{1/6} \sigma$ is the coordinate of LJ minimum, $r_{cut}$ is the cutoff distance, we consider $r_{cut} = \infty$.

The external potential, which acts on fluid particles, is created by pore walls:
\begin{equation}
    V^{ext} = U_{sf}(z) + U_{sf}(H-z)
\end{equation}
where $H$ is the distance between centers of wall molecules on the opposite pore walls, $U_{sf}$ is the solid-fluid interaction potential, which is described by the Steel 10-4-3 potential \cite{steele1974interaction}:
%\begin{equation}
%    U_{sf_i} = 2\pi \rho_s \epsilon_{sf_i} \sigma_{sf_i}^2 \Delta \left[ \left( \frac{\sigma_{sf_i}}{z} \right)^{10} \right. -  \left( \frac{\sigma_{sf_i}}{z} \right)^{4}
%    -  \left. \frac{\sigma_{sf_i}^4}{3 \Delta (0.61\Delta+z)^3}\right]
%\end{equation}
\begin{eqnarray}
    U_{sf} = 2\pi \rho_s \epsilon_{sf} \sigma_{sf}^2 \Delta \left[ \left( \frac{\sigma_{sf}}{z} \right)^{10} \right. -  \left( \frac{\sigma_{sf}}{z} \right)^{4} \nonumber\\
    -  \left. \frac{\sigma_{sf}^4}{3 \Delta (0.61\Delta+z)^3}\right]   
\end{eqnarray}
where $\rho_s$ = 0.114 \AA$^{-3}$, $\Delta$ = 3.35 \AA\, that corresponds to carbon wall. The parameters $\epsilon_{sf}$ and $\sigma_{sf}$ are calculated with the Lorentz\,--\,Berthelot mixing rule with $\epsilon_{s}$ = 28~K and $\sigma_{ss}$ = 3.4 \AA .

The chemical potential can be derived from Helmholtz free energy equations \ref{eq:sum} -- \ref{eq:att} and given by:
\begin{align}
    &\mu = \mu^{id} +\mu^{hs} + \mu^{att}\label{eq:mu_1k}\\
    &\mu^{id} = k_B T \ln{\Lambda^3 \rho}\label{eq:mu_1k_id}\\
    &\mu^{hs} = k_B T \left( \frac{\partial \Phi}{\partial n_3} V + \frac{\partial \Phi}{\partial n_2} S + \frac{\partial \Phi}{\partial n_1} R+ \frac{\partial \Phi}{\partial n_0}\right)\label{eq:mu_1k_hs}\\
    &\mu^{att} = k_B T \rho\int d\vec{r} U^{att}(\vec{r}) \label{eq:mu_1k_att}
\end{align}
with $V = \frac{4}{3}\pi R^3$, $S = \pi R^2$, $R = \sigma/2$ the component particle radius.

The fluid pressure is also derived from Helmholtz free energy equations \ref{eq:sum} -- \ref{eq:att} and gives cDFT EoS:
\begin{align}
    &P = P^{id} +P^{hs} + P^{att}\label{eq:p_2k}\\
    &P^{id} = k_B T \rho  \label{eq:p_1k_id}\\
    &P^{hs} = k_B T \rho  \left(\frac{1 + 2\xi+3\xi^2}{(1-\xi)^2}-1\right)\label{eq:p_2k_hs}\\
    &P^{att} = 0.5 k_B T \rho^2\int d\vec{r} U^{att}(\vec{r})\label{eq:p_1k_att}
\end{align}
with $\xi = \rho V$ is the packing fractions.

\end{document}